\documentclass[english,amssymb,amsmath,noshowpacs,prb,twocolumn]{revtex4-2}
\usepackage[english]{babel}
\usepackage[utf8]{inputenc}
\usepackage{amssymb,amsmath}
\usepackage[colorlinks,pdfpagelabels,pdfstartview = FitH,bookmarksopen = true,bookmarksnumbered = true,linkcolor = black,plainpages = false,hypertexnames = false,citecolor = black,urlcolor=black] {hyperref}
\usepackage{mathrsfs}
\usepackage{mathtools}

\newcommand{\arxivopt}{\\}
\newcommand{\arxiv}{\notag \\}
\newcommand{\arxivcustom}{\hspace{5mm}}
\newcommand{\arxivcustomneu}{\hspace{10mm}}

\begin{document}
	
	\title{Excitonic theory of doping-dependent optical response \\ in atomically thin semiconductors}
	
	\author{Florian Katsch}
	\author{Andreas Knorr}	
	\affiliation{Institut f\"ur Theoretische Physik, Nichtlineare Optik und Quantenelektronik, Technische Universit\"at Berlin, 10623 Berlin, Germany}
	
	\begin{abstract}
		The interaction of optically excited excitons in atomically thin semiconductors with residual doping densities leads to many-body effects which are continuously tunable by external gate voltages.
		Here, we develop a fully microscopic theory to describe the doping-dependent manipulation of the excitonic properties in atomically thin transition metal dichalcogenides.
		In particular, we establish a diagonalization approach for the Schrödinger equation which characterizes the interaction of a virtual exciton with the Fermi sea of dopants.
		Solving this many-body Schrödinger equation provides access to trions as well as a continuum of scattering states.
		The dynamics of coupled excitons, trions, and scattering continua is subsequently described by Heisenberg equations of motion including mean-field contributions and correlation effects due to the interaction of excitons with trions and scattering continuum states.
		Our calculations for optical excitation close to the band edge reveal the influence of doping on the exciton resonances in combination with the simultaneous identification of not only ground- but also excited-state trion resonances.
		\arxivopt
	\end{abstract}
	
	\maketitle
		
\section{Introduction} \label{sec:intro}

	Atomically thin semiconductors combine almost two-dimensional confinement of carriers with weak dielectric screening from the environment which leads to strong Coulomb interaction energies compared to the thermal energies \cite{wang2018colloquium,gies2021atomically}.
	The strong Coulomb attraction between electrons in the conduction bands and holes in the valence bands induces bound electron-hole pairs, called excitons, with binding energies of several hundred meV in monolayer transition metal dichalcogenides (TMDCs) \cite{cheiwchanchamnangij2012quasiparticle,ramasubramaniam2012large,qiu2013optical,chernikov2014exciton,trushin2016optical,trushin2018model,deilmann2019finite}.
	The enhanced Coulomb interaction also provides exceptionally large biexciton \cite{zhang2015excited,hao2017neutral,steinhoff2018biexciton,yong2018biexcitonic,kuhn2019combined,katsch2020theory} and trion binding energies \cite{berkelbach2013theory,mayers2015binding,kylanpaa2015binding,courtade2017charged,kuhn2019tensor}.
	Biexcitons constitute bound Coulomb correlations of two virtual excitons \cite{schafer2013semiconductor,takayama2002t,katsch2020theory2} originating from exciton-exciton interactions.
	Since the biexciton oscillator strength depends on the exciton density, biexcitons can be controlled by the optical excitation power \cite{katsch2020exciton,katsch2020optical}.
	On the other hand, trions or attractive Fermi polarons \cite{mak2013tightly,ross2013electrical}, whose descriptions are equivalent at low doping densities \cite{glazov2020optical}, are bound Coulomb correlations which arise from the interaction of a virtual exciton with the Fermi sea of free conduction band electrons or free valence band holes in doped semiconductors \cite{esser2001theory,glazov2020optical}.
	Consequently, the trion oscillator strength is subject to the doping density and can be regulated by external gate voltages \cite{mak2013tightly,ross2013electrical}.
	\textit{Ab initio} calculations of the trion binding energy are commonly available for monolayer TMDCs \cite{druppel2017diversity,deilmann2017dark,florian2018dielectric,torche2019first,arora2019excited}.
	In contrast, a complete fully microscopic theory of the doping-dependent absorption spectra has not been provided so far:
	Available theoretical studies are based on phenomenological approaches which start from approximate variational exciton and trion states \cite{sidler2017fermi,efimkin2017many,chang2018crossover,efimkin2018exciton,chang2019many,rana2020many,carbone2020microscopic,rana2021many,efimkin2021electron}.
	Most of these studies also rely on phenomenological momentum-independent contact Coulomb potentials \cite{sidler2017fermi,efimkin2017many,efimkin2018exciton,chang2019many} or unscreened two-dimensional Coulomb potentials neglecting the influence of the dielectric environment \cite{efimkin2021electron}.
	In some works, excitons and doping densities are also composed of independent electrons \cite{chang2019many,efimkin2021electron}.
	Moreover, available theoretical studies assume either equal electron and hole masses \cite{efimkin2017many,efimkin2018exciton,chang2019many,rana2020many,efimkin2021electron,rana2021many} or infinite hole masses \cite{chang2018crossover,efimkin2021electron}.
	In particular, no theory is available which can simultaneously describe the doping-dependent absorption of ground-state ($1s$) and excited-state ($2s$) excitons and trions beyond those approximations, which is the novelty of our work.
	In this paper, we establish a theoretical framework for many-particle correlations originating from exciton-electron and exciton-hole interactions in the Heisenberg equation of motion formalism.
	Our theory is restricted to the \textit{linear} optical response and \textit{linear} doping densities, characterized by fully thermalized Fermi distributions, and we treat the band structure in an effective mass approximation.
	In particular, we derive a Schrödinger equation for the interaction between a virtual exciton and the Fermi sea of free electrons or holes that can be exactly solved after separating the relative- and center-of-mass-motion.
	This Schrödinger equation provides bound solutions, referred to as trions which appear energetically below the exciton states, and unbound solutions characterized by exciton-electron or exciton-hole scattering continua which set in at the exciton energy.
	In particular, the Schrödinger equation enables not only access to ground-state ($1s$) trions but also to excited-state ($2s$) trions and corresponding continua.
	The dynamics of excitons, trions, and scattering continua is subsequently determined by coupled Heisenberg equations of motion.
	To relate the developed theory to experimentally accessible observables, we analyze the doping-dependent absorption spectra for monolayer MoSe\textsubscript{2} as a representative atomically thin semiconductor.
	Specifically, we provide a consistent theoretical method to calculate the doping-depend spectra of not only ground-state ($1s$) but also excited-state ($2s$) exciton and trion resonances.
	Since our theory is widely adaptable to semiconductors with tightly bound excitons, we believe that it might provide a basis for further theoretical investigations of doping-dependent excitonic properties.
	This paper is organized as follows:
	In Sec.~\ref{sec:observables}, the microscopic observables including excitons, trions and exciton-electron and exciton-hole continuum states are introduced.
	Afterwards, in Sec.~\ref{sec:factorized-eqs-of-motion}, the dynamics of microscopic observables is described by coupled Heisenberg equations of motion.
	In Sec.~\ref{sec:results}, the doping-dependent absorption spectra are discussed for monolayer MoSe\textsubscript{2} as an exemplary atomically thin semiconductor.
	Finally, in Sec.~\ref{sec:conclusion}, we summarize our results and discuss possible future prospects.

\section{Microscopic Observables} \label{sec:observables}

\begin{figure*}
		\centering
		\includegraphics[width=\textwidth]{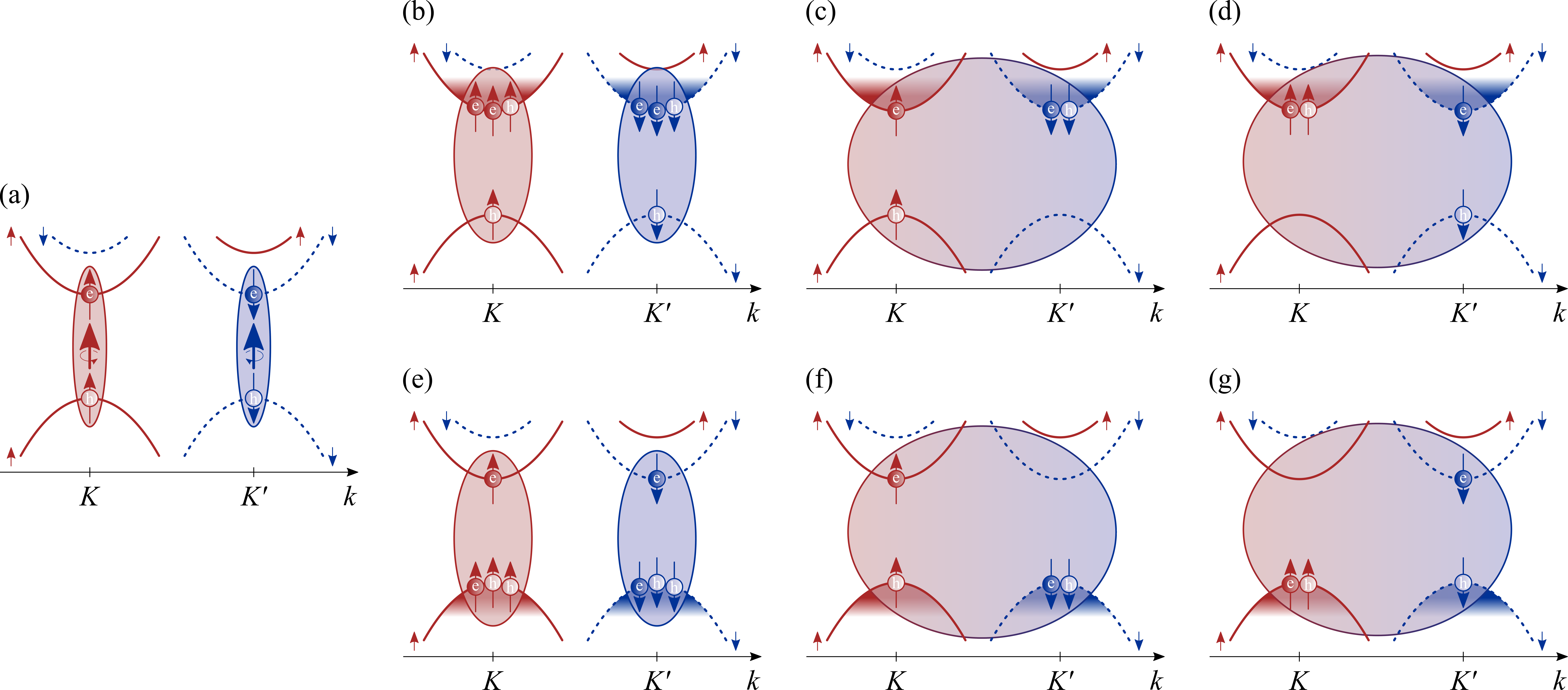}
		\caption{
			Illustration of Coulomb correlations in the simplified band structure for monolayer MoSe\textsubscript{2} near the $K$ and $K'$ high-symmetry points.
			(a)~Optically excited interband transitions are composed of an electron ``$e$'' and a hole ``$h$'' in the same valley.
			(b)~Intravalley and (c),(d)~intervalley electron-density-assisted transitions contribute for electron-doped semiconductors indicated by filled conduction bands.
			In contrast, (e)~intravalley and (f),(g)~intervalley hole-density-assisted transitions contribute in the hole-doping regime illustrated by filled valence bands.
		}
		\label{Bild:Schema}
	\end{figure*}

	In order to describe the exciton dynamics of atomically thin semiconductors with residual doping densities, the infinite hierarchy of coupled Heisenberg equations of motion for many-particle correlations needs to be systematically truncated.
	To this end, we adapt the dynamical truncation scheme \cite{axt1994dynamics,axt1994role}  and restrict to the linear optical response and linear doping densities.
	Maxwell's equations couple the optical light field to the $\sigma_j$~circularly polarized components ($\sigma_j=\sigma_+$, $\sigma_-$) of the macroscopic interband polarization $P\,^{\sigma_j}_{}$ of monolayer TMDCs:
	\begin{equation}
	P\,^{\sigma_j}_{} \left(t\right) = 
	\frac{1}{\mathcal{A}} \sum_{\zeta_1,\textbf{\textit{k}}_1} d\,^{\zeta_1,\sigma_j}_{c,v} \, {\big\langle} c^\dag_{\zeta_1,\textbf{\textit{k}}_1} v^{\phantom{\dagger}}_{\zeta_1,\textbf{\textit{k}}_1} {\big\rangle}\left(t\right) + \text{c.c.} \label{eq:makr-interband-pol} 
	\end{equation}
	$\mathcal{A}$ denotes the two-dimensional normalization area, $\zeta_1 =\{\xi_1,s_1\}$ is a compound index including the valley $\xi_1 = K,$ $K'$ and the spin $s_1 = \ \uparrow$, $\downarrow$, and $\textbf{\textit{k}}_1$ represents the two-dimensional wave vector with respect to the high-symmetry point $\xi_1$.
	The interband dipole transition element $d\,^{\zeta_1,\sigma_j}_{c,v}$ includes the valley-selective circular dichroism of monolayer TMDCs:
	$\sigma_+$~circularly polarized light couples to the $\xi_1 = K$~valley, whereas $\sigma_-$~circularly polarized light couples to the $\xi_1 = K'$~valley \cite{yao2008valley,cao2012valley,zeng2012valley,mak2012control,xiao2012coupled}.
	The interband transitions ${\big\langle} c^\dag_{\zeta_1,\textbf{\textit{k}}_1} v^{\phantom{\dagger}}_{\zeta_1,\textbf{\textit{k}}_1} {\big\rangle}$
	are determined by conduction band creation operators $c^\dag_{\zeta_1,\textbf{\textit{k}}_1}$ and valence band annihilation operators $v_{\zeta_1,\textbf{\textit{k}}_1}$ as illustrated in Fig.~\ref{Bild:Schema}(a).

	In the coherent limit, the Heisenberg equation of motion for the interband transitions ${\big\langle} c^\dag_{\zeta_1,\textbf{\textit{k}}_1} v^{\phantom{\dagger}}_{\zeta_1,\textbf{\textit{k}}_1} {\big\rangle}$ (see Sec.~\ref{sec:factorized-eqs-of-motion}), which determine the interband polarization $P\,^{\sigma_j}_{}$, cf. Eq.~\eqref{eq:makr-interband-pol}, couple to different correlations in the (i)~electron and (ii)~hole doping regimes:
	\begin{itemize}
		\item[(i)] In electron-doped semiconductors the interband transitions ${\big\langle} c^\dag_{\zeta_1,\textbf{\textit{k}}_1} v^{\phantom{\dagger}}_{\zeta_1,\textbf{\textit{k}}_1} {\big\rangle}$ couple to the electron occupations ${\big\langle} c^\dag_{\zeta_1,\textbf{\textit{k}}_1} c^{\phantom{\dagger}}_{\zeta_1,\textbf{\textit{k}}_1} {\big\rangle} = f\,^{\zeta_1}_{e,\textbf{\textit{k}}_1}$, characterized by the residual electron densities $f\,^{\zeta_1}_{e,\textbf{\textit{k}}_1}$, and to the electron-density-assisted transitions ${\big\langle} c^\dag_{\zeta_1,\textbf{\textit{k}}_1+\textbf{\textit{Q}}} v^{\phantom{\dagger}}_{\zeta_1,\textbf{\textit{k}}_1} c^\dag_{\zeta_2,\textbf{\textit{k}}_2-\textbf{\textit{Q}}} c^{\phantom{\dagger}}_{\zeta_2,\textbf{\textit{k}}_2} {\big\rangle}^{c}$, depicted in Figs.~\ref{Bild:Schema}(b)$-$\ref{Bild:Schema}(d):
		\begin{eqnarray}
		& & {\big\langle} c^\dag_{\zeta_1,\textbf{\textit{k}}_1+\textbf{\textit{Q}}} v^{\phantom{\dagger}}_{\zeta_1,\textbf{\textit{k}}_1} c^\dag_{\zeta_2,\textbf{\textit{k}}_2-\textbf{\textit{Q}}} c^{\phantom{\dagger}}_{\zeta_2,\textbf{\textit{k}}_2} {\big\rangle}^{c} \notag \\
		& & = {\big\langle} c^\dag_{\zeta_1,\textbf{\textit{k}}_1+\textbf{\textit{Q}}} v^{\phantom{\dagger}}_{\zeta_1,\textbf{\textit{k}}_1} c^\dag_{\zeta_2,\textbf{\textit{k}}_2-\textbf{\textit{Q}}} c^{\phantom{\dagger}}_{\zeta_2,\textbf{\textit{k}}_2} {\big\rangle} \notag \\
		& & \hspace{3.7mm} - \delta_{\textbf{\textit{Q}},\textbf{0}} \ {\big\langle} c^\dag_{\zeta_1,\textbf{\textit{k}}_1} v^{\phantom{\dagger}}_{\zeta_1,\textbf{\textit{k}}_1} {\big\rangle} f\,^{\zeta_2}_{e,\textbf{\textit{k}}_2} \notag \\
		& & \hspace{3.7mm} + \delta_{\zeta_1,\zeta_2} \ \delta_{\textbf{\textit{k}}_1+\textbf{\textit{Q}},\textbf{\textit{k}}_2} \ {\big\langle} c^\dag_{\zeta_1,\textbf{\textit{k}}_1} v^{\phantom{\dagger}}_{\zeta_1,\textbf{\textit{k}}_1} {\big\rangle} f\,^{\zeta_1}_{e,\textbf{\textit{k}}_1+\textbf{\textit{Q}}} ,
		\label{eq:fat-e}
		\end{eqnarray}
		
		\item[(ii)] In hole-doped semiconductors the interband transitions ${\big\langle} c^\dag_{\zeta_1,\textbf{\textit{k}}_1} v^{\phantom{\dagger}}_{\zeta_1,\textbf{\textit{k}}_1} {\big\rangle}$ couple to the hole occupations ${\big\langle} v^{\phantom{\dagger}}_{\zeta_1,\textbf{\textit{k}}_1} v^\dag_{\zeta_1,\textbf{\textit{k}}_1} {\big\rangle}=f\,^{\zeta_1}_{h,\textbf{\textit{k}}_1}$, determined by the residual hole densities $f\,^{\zeta_1}_{h,\textbf{\textit{k}}_1}$, and to the hole-density-assisted transitions ${\big\langle} c^\dag_{\zeta_1,\textbf{\textit{k}}_1+\textbf{\textit{Q}}} v^{\phantom{\dagger}}_{\zeta_1,\textbf{\textit{k}}_1} v^{\phantom{\dagger}}_{\zeta_2,\textbf{\textit{k}}_2+\textbf{\textit{Q}}} v^\dag_{\zeta_2,\textbf{\textit{k}}_2} {\big\rangle}^{c}$, illustrated in Figs.~\ref{Bild:Schema}(e)$-$\ref{Bild:Schema}(g):
		\begin{eqnarray}
		& & {\big\langle} c^\dag_{\zeta_1,\textbf{\textit{k}}_1+\textbf{\textit{Q}}} v^{\phantom{\dagger}}_{\zeta_1,\textbf{\textit{k}}_1} v^{\phantom{\dagger}}_{\zeta_2,\textbf{\textit{k}}_2+\textbf{\textit{Q}}} v^\dag_{\zeta_2,\textbf{\textit{k}}_2} {\big\rangle}^{c} \notag \\
		& & = {\big\langle} c^\dag_{\zeta_1,\textbf{\textit{k}}_1+\textbf{\textit{Q}}} v^{\phantom{\dagger}}_{\zeta_1,\textbf{\textit{k}}_1} v^{\phantom{\dagger}}_{\zeta_2,\textbf{\textit{k}}_2+\textbf{\textit{Q}}} v^\dag_{\zeta_2,\textbf{\textit{k}}_2} {\big\rangle} \notag \\
		& & \hspace{3.7mm} - \delta_{\textbf{\textit{Q}},\textbf{0}} \ {\big\langle} c^\dag_{\zeta_1,\textbf{\textit{k}}_1} v^{\phantom{\dagger}}_{\zeta_1,\textbf{\textit{k}}_1} {\big\rangle} f\,^{\zeta_2}_{h,\textbf{\textit{k}}_2} \notag \\
		& & \hspace{3.7mm} + \delta_{\zeta_1,\zeta_2} \ \delta_{\textbf{\textit{k}}_1,\textbf{\textit{k}}_2} \ {\big\langle} c^\dag_{\zeta_1,\textbf{\textit{k}}_1+\textbf{\textit{Q}}} v^{\phantom{\dagger}}_{\zeta_1,\textbf{\textit{k}}_1+\textbf{\textit{Q}}} {\big\rangle} f\,^{\zeta_1}_{h,\textbf{\textit{k}}_1} . \label{eq:fat-h}
		\end{eqnarray}		
	\end{itemize}
	Here, ${\big\langle} \cdot {\big\rangle}^{c}$ denotes the correlated part \cite{fricke1996transport}.
	The truncation approach is valid in the limit of linear optical excitation.
	Optically excited electron densities accompanied by simultaneously excited hole densities, due to assigning a hole to every optically generated electron in the conduction band \cite{katsch2018theory}, were neglected.
	The residual electron ``$e$'' or hole ``$h$'' densities $f\,^{\zeta}_{e/h,\textbf{\textit{k}}}$ in the atomically thin semiconductor are approximated by fully thermalized Fermi distributions:
	\begin{align}
	f\,^{\zeta}_{e/h,\textbf{\textit{k}}} = 
	\frac{1}{\exp\big[\frac{1}{k_B T}\big( \varepsilon^{\zeta}_{e/h,\textbf{\textit{k}}}-\mu_{e/h} \big)\big] +1} \label{eq:Fermi-dist} .
	\end{align}
	$k_B$ denotes the Boltzmann constant, $T$ is the temperature, and $\mu_{e/h}$ represents the chemical potential.
	The electron and hole dispersion $\varepsilon^{\zeta}_{e/h,\textbf{\textit{k}}} = \varepsilon\,^{\zeta}_g/2+{\hbar^2\textbf{\textit{k}}^2}/{(2m_{e/h})}$ are treated in an effective mass approximation and involve the band gap energy $\varepsilon\,^{\zeta}_g$ between conduction and valence band edges as well as the effective electron or hole mass $m_{e/h}$.
	The total electron ``$e$'' and hole ``$h$'' densities $N_{e/h}$ are given by:
	\begin{equation}
	N_{e/h} = \frac{1}{\mathcal{A}} \sum_{\zeta,\textbf{\textit{k}}} f\,^{\zeta}_{e/h,\textbf{\textit{k}}} .
	\end{equation}
	Since we focus on the lowest conduction bands and topmost valence bands, the effective masses $m_{e}$ and $m_{h}$ were chosen to be independent of the compound index $\zeta$.
	Of course, an extension to valley- and spin-dependent effective masses is possible.
	The treatment of optically excited interband transitions as excitons is described in Sec.~\ref{subsec:x-obs}.
	Trions and exciton-electron or exciton-hole continua necessary to describe the linear optical response of electron- or hole-doped semiconductors are introduced in Sec.~\ref{subsec:neg-obs}.

\subsection{Excitons} \label{subsec:x-obs}

	Interband transitions ${\big\langle} c^\dag_{\zeta_1,\textbf{\textit{k}}_1} v^{\phantom{\dagger}}_{\zeta_1,\textbf{\textit{k}}_1} {\big\rangle}$, as depicted in Fig.~\ref{Bild:Schema}(a), determine the interband polarization $P\,^{\sigma_j}_{}$ according to Eq.~\eqref{eq:makr-interband-pol}, and are subsequently treated by the Wannier equation \cite{kira2006many}:
	\begin{equation}
	\frac{\hbar^2 \textbf{\textit{k}}_1^2}{2 \mu} \varphi^R\,^{\zeta_1}_{\nu,\textbf{\textit{k}}_1} - \sum_{\textbf{\textit{k}}_2} W_{\textbf{\textit{k}}_1-\textbf{\textit{k}}_2} \ \varphi^R\,^{\zeta_1}_{\nu,\textbf{\textit{k}}_2}  = \left(\epsilon\,^{\zeta_1}_{x,\nu} - \varepsilon\,^{\zeta_1}_g \right) \varphi^R\,^{\zeta_1}_{\nu,\textbf{\textit{k}}_1} \label{eq:Wannier-Gl} .
	\end{equation}
	Instead of introducing doping-dependent filling factors in the Wannier equation \cite{steinhoff2017exciton}, we include these contributions later in the equations of motion for exciton transitions.
	Solving the Wannier equation, Eq.~\eqref{eq:Wannier-Gl}, provides a complete set of wave functions~$ \varphi^R\,^{\zeta_1}_{\nu,\textbf{\textit{k}}_1}$ and corresponding energies $\epsilon\,^{\zeta_1}_{x,\nu}$ indicated by the quantum number $\nu$.
	The Wannier equation has both left- $\varphi^L\,^{\zeta_1}_{\nu,\textbf{\textit{k}}_1}$ and right-handed $\varphi^R\,^{\zeta_1}_{\nu,\textbf{\textit{k}}_1}$ solutions which are normalized as follows \cite{kira2006many}:
	\begin{equation}
		\frac{1}{\mathcal{A}} \sum_{\textbf{\textit{k}}_1} \varphi^L\,^{\zeta_1}_{\nu_1,\textbf{\textit{k}}_1} \ \varphi^R\,^{\zeta_1}_{\nu_2,\textbf{\textit{k}}_1} = \delta_{\nu_1,\nu_2} . \label{eq:normierung-x}
	\end{equation}
	Since Eq.~\eqref{eq:Wannier-Gl} is Hermitian, its left- and right-handed solutions satisfy: $\varphi^L\,^{\zeta_1}_{\nu,\textbf{\textit{k}}_1} = \big(\varphi^R\,^{\zeta_1}_{\nu,\textbf{\textit{k}}_1}\big)^*$ \cite{kira2006many}.
	The Wannier equation, Eq.~\eqref{eq:Wannier-Gl}, depends on the reduced mass $\mu = m_{e}m_{h}/(m_{e}+m_{h})$ which is defined with respect to the effective electron and hole masses $m_{e}$ and $m_{h}$.
	The screened Coulomb potential $W_{\textbf{\textit{k}}} = V_{\textbf{\textit{k}}}/\varepsilon_{\textbf{\textit{k}}}$ involves the bare Coulomb potential $V_{\textbf{\textit{k}}}$ and the screening function $\varepsilon_{\textbf{\textit{k}}}$ given in Appendix~\ref{app:screen}.
	The screened Coulomb potential is obtained from solving Poison's equation for the following van der Waals heterostructure: environment/air/atomically thin semiconductor/air/environment \cite{florian2018dielectric,steinhoff2020dynamical}.
	The small air gaps account for naturally occurring but non-vanishing interlayer distances between the atomically thin semiconductor and its dielectric environment characterized by the dielectric constant $\varepsilon_e$ \cite{rooney2017observing}.
	In the following, we expand the interband transitions ${\big\langle} c^\dag_{\zeta_1,\textbf{\textit{k}}_1} v^{\phantom{\dagger}}_{\zeta_1,\textbf{\textit{k}}_1} {\big\rangle}$ in terms of the complete set of exciton wave functions $\varphi^R\,^{\zeta_1}_{\nu,\textbf{\textit{k}}_1}$ and associated expansion coefficients represented by the exciton transitions $P\,^{\zeta_1}_{\nu}$:
	\begin{equation}
	{\big\langle} c^\dag_{\zeta_1,\textbf{\textit{k}}_1} v^{\phantom{\dagger}}_{\zeta_1,\textbf{\textit{k}}_1} {\big\rangle}{(t)}
	=  \sum_{\nu} \varphi^R\,^{\zeta_1}_{\nu,\textbf{\textit{k}}_1} \ P\,^{\zeta_1}_{\nu}{(t)} . \label{eq:definition-exciton}
	\end{equation}
	As a result, the excitonic expansion of the macroscopic interband polarization $P\,^{\sigma_j}_{}$ defined in Eq.~\eqref{eq:makr-interband-pol} becomes:
	\begin{equation}
	P\,^{\sigma_j}_{} \left(t\right) = 
	\frac{1}{\mathcal{A}} \sum_{\zeta_1,\nu,\textbf{\textit{k}}_1} \varphi^R\,^{\zeta_1}_{\nu,\textbf{\textit{k}}_1} \, d\,^{\zeta_1,\sigma_j}_{c,v\vphantom{\textbf{\textit{k}}_1}} \, P\,^{\zeta_1}_{\nu}\left(t\right) + \text{c.c.} \label{eq:makr-interband-pol1} 
	\end{equation}

\subsection{Trions and Exciton-Electron/Hole Continua} \label{subsec:neg-obs}

	The treatment of electron-density-assisted transitions ${\big\langle} c^\dag_{\zeta_1,\textbf{\textit{k}}_1+\textbf{\textit{Q}}} v^{\phantom{\dagger}}_{\zeta_1,\textbf{\textit{k}}_1} c^\dag_{\zeta_2,\textbf{\textit{k}}_2-\textbf{\textit{Q}}} c^{\phantom{\dagger}}_{\zeta_2,\textbf{\textit{k}}_2} {\big\rangle}^{c}$, illustrated in Figs.~\ref{Bild:Schema}(b)$-$\ref{Bild:Schema}(d), and hole-density-assisted transitions ${\big\langle} c^\dag_{\zeta_1,\textbf{\textit{k}}_1+\textbf{\textit{Q}}} v^{\phantom{\dagger}}_{\zeta_1,\textbf{\textit{k}}_1} v^{\phantom{\dagger}}_{\zeta_2,\textbf{\textit{k}}_2+\textbf{\textit{Q}}} v^\dag_{\zeta_2,\textbf{\textit{k}}_2} {\big\rangle}^{c}$, depicted in Figs.~\ref{Bild:Schema}(e)$-$\ref{Bild:Schema}(g), is derived in the following.
	The approach is based on the projection onto excitonic wave functions and electron or hole densities.
	Due to the anticommutation of the two electron creation operators, the  electron-density-assisted transitions ${\big\langle} c^\dag_{\zeta_1,\textbf{\textit{k}}_1+\textbf{\textit{Q}}} v^{\phantom{\dagger}}_{\zeta_2,\textbf{\textit{k}}_1} c^\dag_{\zeta_3,\textbf{\textit{k}}_2-\textbf{\textit{Q}}} c^{\phantom{\dagger}}_{\zeta_4,\textbf{\textit{k}}_2} {\big\rangle}^{c}$ satisfy:
	\begin{eqnarray}
	& & {\big\langle} c^\dag_{\zeta_1,\textbf{\textit{k}}_1+\textbf{\textit{Q}}} v^{\phantom{\dagger}}_{\zeta_2,\textbf{\textit{k}}_1} c^\dag_{\zeta_3,\textbf{\textit{k}}_2-\textbf{\textit{Q}}} c^{\phantom{\dagger}}_{\zeta_4,\textbf{\textit{k}}_2} {\big\rangle}^{c} \notag \\
	& & = - {\big\langle} c^\dag_{\zeta_3,\textbf{\textit{k}}_2-\textbf{\textit{Q}}} v^{\phantom{\dagger}}_{\zeta_2,\textbf{\textit{k}}_1} c^\dag_{\zeta_1,\textbf{\textit{k}}_1+\textbf{\textit{Q}}} c^{\phantom{\dagger}}_{\zeta_4,\textbf{\textit{k}}_2} {\big\rangle}^{c} .\label{eq:antisymmetry}
	\end{eqnarray}	
	This antisymmetry property is fulfilled by the following expansion into products of exciton wave functions \smash[b]{\smash[t]{$\varphi^R\,^{\zeta_2}_{\nu,\textbf{\textit{k}}_1}$}}, electron densities \smash[b]{\smash[t]{$f\,^{\zeta_4}_{e,\textbf{\textit{k}}_2}$}}, and associated expansion coefficients \smash[b]{\smash[t]{$\mathcal{T}\,^{\zeta_1,\zeta_2,\zeta_3,\zeta_4}_{x\text{-}e,\nu,\textbf{\textit{Q}}}$}}:
	\begin{eqnarray}
	& & {\big\langle} c^\dag_{\zeta_1,\textbf{\textit{k}}_1+\alpha_{x\text{-}e}\textbf{\textit{k}}_2+\alpha_{x}\textbf{\textit{Q}}} v^{\phantom{\dagger}}_{\zeta_2,\textbf{\textit{k}}_1-\beta_{x\text{-}e}\textbf{\textit{k}}_2-\beta_{x}\textbf{\textit{Q}}} c^\dag_{\zeta_3,\alpha_{x\text{-}e}\textbf{\textit{k}}_2-\textbf{\textit{Q}}} c^{\phantom{\dagger}}_{\zeta_4,\textbf{\textit{k}}_2} {\big\rangle}^{c} \notag \\
	& & = \sum_{\nu} \Big( \varphi^R\,^{\zeta_2}_{\nu,\textbf{\textit{k}}_1} \ \mathcal{T}\,^{\zeta_1,\zeta_2,\zeta_3,\zeta_4}_{x\text{-}e,\nu,\textbf{\textit{Q}}} \notag \\
	& & \hspace{11.3mm}- \varphi^R\,^{\zeta_2}_{\nu,\alpha_{x}\textbf{\textit{k}}_1+[\alpha_{x}^2-1]\textbf{\textit{Q}}} \ \mathcal{T}\,^{\zeta_3,\zeta_2,\zeta_1,\zeta_4}_{x\text{-}e,\nu,-\textbf{\textit{k}}_1-\alpha_{x}\textbf{\textit{Q}}} \Big) f\,^{\zeta_4}_{e,\textbf{\textit{k}}_2} \label{eq:four-particle-corr} .
	\end{eqnarray}
	This expansion is enabled by the conveniently chosen wave vectors which separate the relative- and center-of-mass-dependent parts of the dynamics, see Appendix~\ref{App:derivation}:
	The relative motion of the electron-hole pair is described by excitonic wave functions obtained by solving the Wannier equation, Eq.~\eqref{eq:Wannier-Gl}.
	The relative motion of the second electron creation and annihilation operator is characterized by the distribution function of the residual electron density.
	Even though, we use temperature-dependent Fermi distributions, our theory can be also applied to a different distribution. The center-of-mass motion of the electron-density-assisted transitions is represented by the expansion coefficients which are determined in the following.
	The second term on the right-hand side of Eq.~\eqref{eq:four-particle-corr} ensures that the antisymmetry property of electron-density-assisted transitions is fulfilled, Eq.~\eqref{eq:antisymmetry}, and describes the corresponding expansion for exchanged electron creation operators.
	The two terms on the right-hand side of Eq.~\eqref{eq:four-particle-corr} account for the two possibilities to match every of the two electron creation operators with the valence band annihilation operator.
	The chosen wave vector coordinates involve the ratios of effective masses $\alpha_{x}$, $\beta_{x}$, $\alpha_{x\text{-}e}$, $\beta_{x\text{-}e}$, $\alpha_{x\text{-}h}$, and $\beta_{x\text{-}h}$ defined by:
	\begin{eqnarray}
	\alpha_{x} & = & \frac{m_{e}}{m_{e}+m_{h}} , \hspace{6.3mm} \beta_{x} = \frac{m_{h}}{m_{e}+m_{h}} , \label{eq:alpha-beta} \\
	\alpha_{x\text{-}e} & = & \frac{m_{e}}{2m_{e}+m_{h}} , \hspace{2.3mm} \beta_{x\text{-}e} = \frac{m_{h}}{2m_{e}+m_{h}} , \label{eq:alpha-beta-e} \\
	\alpha_{x\text{-}h} & = & \frac{m_{e}}{m_{e}+2m_{h}} , \hspace{2mm} \beta_{x\text{-}h} = \frac{m_{h}}{m_{e}+2m_{h}} .
	\end{eqnarray}
	Next, it will prove beneficial to introduce symmetric ``$+$'' and antisymmetric ``$-$'' linear combinations of the correlation function defined in Eq.~\eqref{eq:four-particle-corr}:
	\begin{widetext}\vspace{-2.7mm}
	\begin{eqnarray}
		& & \frac{1}{2}\bigg({\big\langle} c^\dag_{\zeta_1,\textbf{\textit{k}}_1+\alpha_{x\text{-}e}\textbf{\textit{k}}_2+\alpha_{x}\textbf{\textit{Q}}} v^{\phantom{\dagger}}_{\zeta_1,\textbf{\textit{k}}_1-\beta_{x\text{-}e}\textbf{\textit{k}}_2-\beta_{x}\textbf{\textit{Q}}} c^\dag_{\zeta_2,\alpha_{x\text{-}e}\textbf{\textit{k}}_2-\textbf{\textit{Q}}} c^{\phantom{\dagger}}_{\zeta_2,\textbf{\textit{k}}_2} {\big\rangle}^{c}
		\pm {\big\langle} c^\dag_{\zeta_2,\textbf{\textit{k}}_1+\alpha_{x\text{-}e}\textbf{\textit{k}}_2+\alpha_{x}\textbf{\textit{Q}}} v^{\phantom{\dagger}}_{\zeta_1,\textbf{\textit{k}}_1-\beta_{x\text{-}e}\textbf{\textit{k}}_2-\beta_{x}\textbf{\textit{Q}}} c^\dag_{\zeta_1,\alpha_{x\text{-}e}\textbf{\textit{k}}_2-\textbf{\textit{Q}}} c^{\phantom{\dagger}}_{\zeta_2,\textbf{\textit{k}}_2} {\big\rangle}^{c}\bigg) \notag \\
		& & = \sum_{\nu} \left(\varphi^R\,^{\zeta_1}_{\nu,\textbf{\textit{k}}_1} \ \hat{T}\,^{\zeta_1,\zeta_2}_{x\text{-}e,\pm,\nu,\textbf{\textit{Q}}} \mp \varphi^R\,^{\zeta_1}_{\nu,\alpha_{x}\textbf{\textit{k}}_1+[\alpha_{x}^2-1]\textbf{\textit{Q}}} \ \hat{T}\,^{\zeta_1,\zeta_2}_{x\text{-}e,\pm,\nu,-\textbf{\textit{k}}_1-\alpha_{x}\textbf{\textit{Q}}}\right) f\,^{\zeta_2}_{e,\textbf{\textit{k}}_2} . \label{eq:lin-comb}
	\end{eqnarray}
	The ``$+$'' (triplet) configuration describes states which are symmetric under exchange of the two electron creation operators, whereas the ``$-$'' (singlet) configuration is antisymmetric with respect to the interchange of the two electrons.
	The new expansion coefficients~$\hat{T}\,^{\zeta_1,\zeta_2}_{x\text{-}e,\pm,\nu,\textbf{\textit{Q}}}$ on the right-hand side of Eq.~\eqref{eq:lin-comb} are defined by:
	$\hat{T}\,^{\zeta_1,\zeta_2}_{x\text{-}e,\pm,\nu,\textbf{\textit{Q}}} = \frac{1}{2} \big(\mathcal{T}\,^{\zeta_1,\zeta_1,\zeta_2,\zeta_2}_{x\text{-}e,\nu,\textbf{\textit{Q}}} \pm \mathcal{T}\,^{\zeta_2,\zeta_1,\zeta_1,\zeta_2}_{x\text{-}e,\nu,\textbf{\textit{Q}}}\big) $.
	An analogous expansion of the hole-density-assisted transitions ${\big\langle} c^\dag_{\zeta_1,\textbf{\textit{k}}_1+\textbf{\textit{Q}}} v^{\phantom{\dagger}}_{\zeta_1,\textbf{\textit{k}}_1} v^{\phantom{\dagger}}_{\zeta_2,\textbf{\textit{k}}_2+\textbf{\textit{Q}}} v^\dag_{\zeta_2,\textbf{\textit{k}}_2} {\big\rangle}^{c}$ is derived in Appendix~\ref{subsec:pos-obs}.
	The Coulomb correlations are treated by solving an associated Schrödinger equation:
	\begin{eqnarray}
	& & \frac{\hbar^2\textbf{\textit{Q}}_1^2}{2}\left(\frac{1}{m_{e}+m_{h}}+\frac{1}{m_{e/h}}\right) \psi^R\,^{\zeta_1,\zeta_2}_{x\text{-}e/h,\pm,\mu,\nu_1,\textbf{\textit{Q}}_1}
	+ \sum_{\nu_2,\textbf{\textit{Q}}_2} \big(S\,^{\zeta_1}_{x\text{-}e/h,\pm}\big)^{-1}_{\nu_1,\nu_2,\textbf{\textit{Q}}_1,\textbf{\textit{Q}}_2} \sum_{\nu_3,\textbf{\textit{Q}}_3} {W}\,^{\zeta_1}_{x\text{-}e/h,\pm,\nu_2,\nu_3,\textbf{\textit{Q}}_2,\textbf{\textit{Q}}_3} \ \psi^R\,^{\zeta_1,\zeta_2}_{x\text{-}e/h,\pm,\mu,\nu_3,\textbf{\textit{Q}}_3} \notag \\
	& & = \left(\epsilon\,^{\zeta_1,\zeta_2}_{x\text{-}e/h,\pm,\mu} - \epsilon\,^{\zeta_1}_{x,\nu_1}\right) \ \psi^R\,^{\zeta_1,\zeta_2}_{x\text{-}e/h,\pm,\mu,\nu_1,\textbf{\textit{Q}}_1} \label{eq:Bi-X} .
	\end{eqnarray}	
\vspace{-4.7mm}\end{widetext}
	Equation~\eqref{eq:Bi-X} provides a complete set of wave functions \smash[b]{\smash[t]{$\psi^R\,^{\zeta_1,\zeta_2}_{x\text{-}e/h,\pm,\mu,\nu_1,\textbf{\textit{Q}}_1}$}} with real-valued energies \smash[b]{\smash[t]{$\epsilon\,^{\zeta_1,\zeta_2}_{x\text{-}e/h,\pm,\mu}$}} indicated by the quantum number $\mu$ with respect to the exciton energy \smash[b]{\smash[t]{$\epsilon\,^{\zeta_1}_{x,\nu_1}$}}.
	Compared to the standard Schrödinger equation for electron/hole-density-assisted transitions \cite{stebe1998optical,esser2000photoluminescence,sergeev2001triplet,esser2001theory}, as derived in Appendix~\ref{App:derivation}, Eq.~\eqref{eq:Bi-X} describes only the center-of-mass motion depending on the wave vector $\textbf{\textit{Q}}$ of the electron/hole-density-assisted transitions.
	According to Eq.~\eqref{eq:four-particle-corr} the full wave function of the electron/hole-density-assisted transitions also includes the excitonic wave function and the distribution function of the residual electron density in addition to the wave function \smash[b]{\smash[t]{$\psi^R\,^{\zeta_1,\zeta_2}_{x\text{-}e/h,\pm,\mu,\nu_1,\textbf{\textit{Q}}_1}$}} and therefore depends on three wave vectors.
	Thus, Eq.~\eqref{eq:four-particle-corr} does not imply a treatment of electron/hole-density-assisted transitions as a rigid exciton attached to the Fermi sea of dopants.
	Instead, the possibility to separate the center-of-mass motion results without further approximations originates from the combination of the conveniently chosen wave vectors and the treatment of symmetric and antisymmetric linear combinations, which facilitate the separation ansatz, Eq.~\eqref{eq:four-particle-corr}.
	A detailed derivation can be found in Appendix~\ref{App:derivation}.
	Compared to a previously developed description \cite{rana2020many}, our derived Schrödinger equation for electron/hole-density-assisted transitions, Eq.~\eqref{eq:Bi-X}, allows for an exact diagonalization to obtain trions and continuum states.
	However, our description is restricted to linear doping densities, which characterize a lower doping regime compared to Ref.~\cite{rana2020many}.

	Since Eq.~\eqref{eq:Bi-X} is non-Hermitian, there are both left- $\psi^L\,^{\zeta_1,\zeta_2}_{x\text{-}e/h,\pm,\mu_1,\nu_1,\textbf{\textit{Q}}_1}$ and right-handed $\psi^R\,^{\zeta_1,\zeta_2}_{x\text{-}e/h,\pm,\mu_2,\nu_1,\textbf{\textit{Q}}_1}$ solutions which are normalized as follows:
	\begin{equation}
	\sum_{\nu_1,\textbf{\textit{Q}}_1} \psi^L\,^{\zeta_1,\zeta_2}_{x\text{-}e/h,\pm,\mu_1,\nu_1,\textbf{\textit{Q}}_1} \ \psi^R\,^{\zeta_1,\zeta_2}_{x\text{-}e/h,\pm,\mu_2,\nu_1,\textbf{\textit{Q}}_1} = \delta_{\mu_1,\mu_2} .
	\end{equation}
	The first term on the left-hand side of Eq.~\eqref{eq:Bi-X} characterizes diagonal contributions which represent the relative motion determined by the wave vector $\textbf{\textit{Q}}_1$.
	Coulomb interactions are described by the second term on the left-hand side of Eq.~\eqref{eq:Bi-X} and constitute both diagonal and non-diagonal contributions.
	The appearance of the inverse matrix of $S\,^{\zeta_1}_{x\text{-}e/h,\pm,\nu_1,\nu_2,\textbf{\textit{Q}}_1,\textbf{\textit{Q}}_2}$, defined by Eqs.~\eqref{eq:S-Matrix} and \eqref{eq:S-Matrix2}, traces back to the treatment of symmetric and antisymmetric linear combinations of the Coulomb correlations in Eq.~\eqref{eq:lin-comb}.
	The Coulomb interaction kernel $W\,^{\zeta_1}_{x\text{-}e/h,\pm,\nu_1,\nu_2,\textbf{\textit{Q}}_1,\textbf{\textit{Q}}_2}$ is given in Eqs.~\eqref{eq:Coulomb-M-E} and \eqref{eq:Coulomb-M-E-2}.

	\begin{figure}
		\centering
		\includegraphics[width=1\columnwidth]{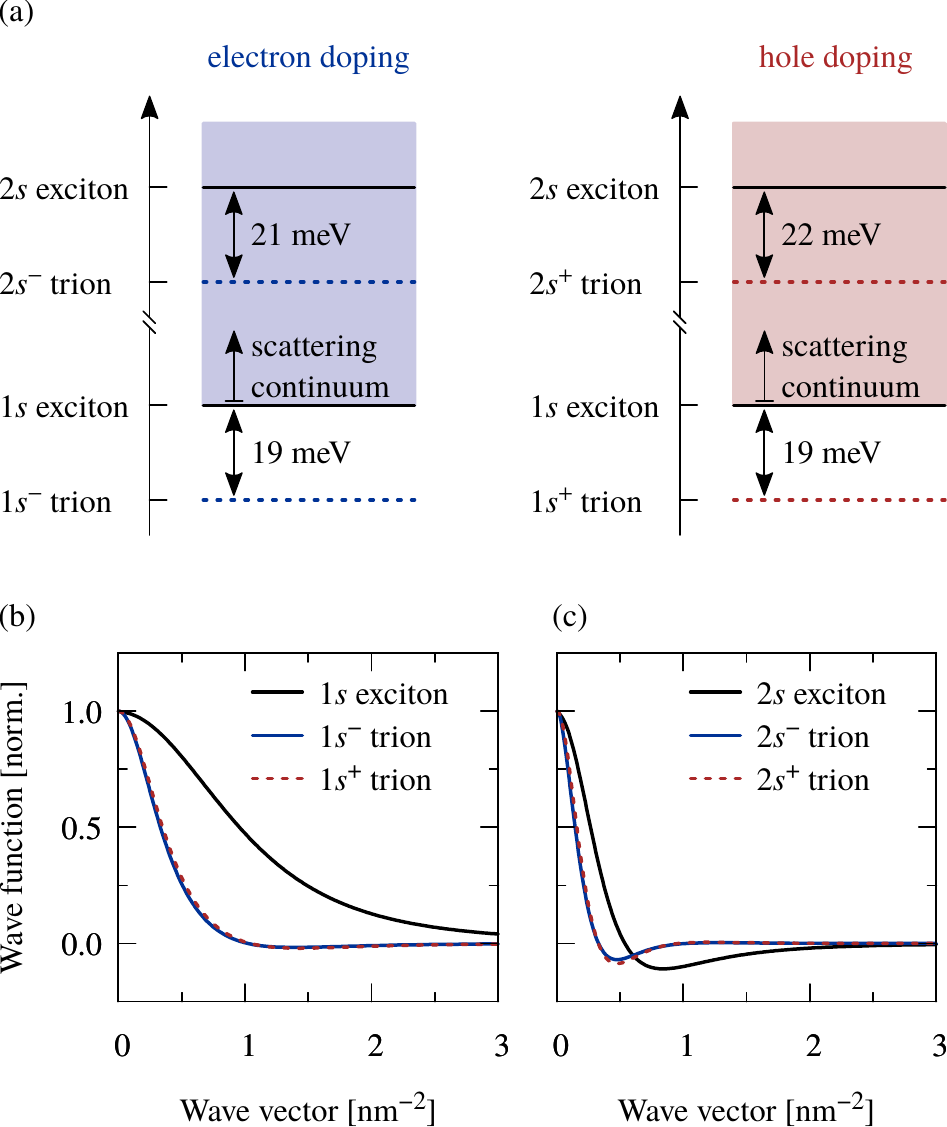}
		\caption{
			(a)~Binding energies of negatively/positively charged ground-state $1s^{-/+}$ and excited-state $2s^{-/+}$ trions with respect to charge neutral $1s$ and $2s$ excitons in an electron and hole doped monolayer MoSe\textsubscript{2} encapsulated in hexagonal BN.
			The exciton-electron/hole scattering continua set in at the $1s$ exciton energy and are illustrated as shaded areas.
			(b)~Normalized radial parts of the wave functions of ground-state $1s$ excitons and $1s^{-/+}$ trions for monolayer MoSe\textsubscript{2} encapsulated in hexagonal BN.
			(c)~Corresponding wave functions of excited-state $2s$ excitons and $2s^{-/+}$ trions.
		}
		\label{Bild:wellenfunktionen}
	\end{figure}

	Solving Eq.~\eqref{eq:Bi-X} for the \mbox{``$+$''} configuration provides a continuum of energetically dense states with state energies $\epsilon\,^{\zeta_1,\zeta_2}_{x\text{-}e/h,+,\mu} \ge \epsilon\,^{\zeta_1}_{x,1s}$ which are unbound with respect to the $\nu = 1s$ exciton energy $\epsilon\,^{\zeta_1}_{x,1s}$.
	These dense states are referred to as the exciton-electron/hole continuum and illustrated as shaded areas in Fig.~\ref{Bild:wellenfunktionen}(a).
	In contrast, the solutions of Eq.~\eqref{eq:Bi-X} for the \mbox{``$-$''} configuration also comprise bound states, called trions $\mu=t$, with energies $\epsilon\,^{\zeta_1,\zeta_2}_{x\text{-}e/h,-,t}<\epsilon\,^{\zeta_1}_{x,1s}$ smaller than the $1s$ exciton energy.
	Trions are depicted as dashed lines in Fig.~\ref{Bild:wellenfunktionen}(a).
	Note that even though we use the term trion in the following, we treat trions according to the Fermi-polaron picture as four-particle complexes.
	Of course the \mbox{``$-$''} configuration also provides exciton-electron/hole continuum states $\mu\neq t$ characterizing many dense states with energies $\epsilon\,^{\zeta_1,\zeta_2}_{x\text{-}e/h,-,\mu\neq t} \ge \epsilon\,^{\zeta_1}_{x,1s}$.
	Since the \mbox{``$-$''} configuration is trivially zero for identical compound valley-spin indices $\zeta_1 = \zeta_2$, intravalley trions with same spins are naturally excluded, see Eq.~\eqref{eq:lin-comb}.
	Note that even though we refer to the $\mu=t$ states as trions they can also be understood as attractive Fermi polarons \cite{sidler2017fermi,efimkin2017many} because their descriptions are equivalent at low doping densities \cite{glazov2020optical}.
	To compare our calculations to previous theoretical predictions, we first discuss freestanding ($\varepsilon_e=1$) monolayer MoSe\textsubscript{2}.
	Solving Eq.~\eqref{eq:Bi-X} locates the negatively charged $1s^-$ trion 28~meV below the $1s$ exciton.
	This value is in agreement with former theoretical calculations obtaining 21~meV to 35~meV for the $1s^-$ trion binding energy \cite{berkelbach2013theory,mayers2015binding,kylanpaa2015binding,szyniszewski2017binding,zhang2015excited,kidd2016binding,van2017excitons,kezerashvili2017trion,mostaani2017diffusion,van2018excitons,florian2018dielectric,tempelaar2019many,fey2020theory}.
	On the other hand, solving Eq.~\eqref{eq:Bi-X} locates the positively charged $1s^+$~trion 27~meV below the $1s$ exciton, which is close to the theoretical predictions of 28~meV to 34~meV \cite{kylanpaa2015binding,mostaani2017diffusion,florian2018dielectric}.
	However, atomically thin semiconductors are typically embedded in a dielectric environment and we will subsequently focus on monolayer MoSe$_2$ encapsulated in hexagonal BN ($\varepsilon_e=4.5$).
	The environment results in enhanced dielectric screening and decreases the trion binding energies with rising dielectric constant $\varepsilon_e$ of the environment \cite{kylanpaa2015binding}.
	As a result, encapsulation of monolayer MoSe\textsubscript{2} in hexagonal BN reduces the binding energies of $1s^-$ and $1s^+$ trions to 19~meV as illustrated in Fig.~\ref{Bild:wellenfunktionen}(a).
	In addition to the ground-state $1s^-$ and $1s^+$ trions, Eq.~\eqref{eq:Bi-X} also provides excited-state $2s^-$ and $2s^+$ trions depicted in Fig.~\ref{Bild:wellenfunktionen}(a) which appear 21~meV and 22~meV below the $2s$ excitons, respectively.
	The outcome of excited-state $2s^{-/+}$ trions is in agreement with recent \textit{ab initio} calculations \cite{arora2019excited}.
	The normalized radial parts of wave functions $\psi^R\,^{\{K,\uparrow\},\{K',\downarrow\}}_{x\text{-}e/h,-,1s^{-/+},1s,\textbf{\textit{Q}}}$ for $1s^{-/+}$ trions obtained for monolayer MoSe\textsubscript{2} encapsulated in hexagonal BN are plotted as blue solid and red dashed lines in Fig.~\ref{Bild:wellenfunktionen}(b).
	The wave functions $\psi^R\,^{\{K,\uparrow\},\{K',\downarrow\}}_{x\text{-}e/h,-,1s^{-/+},1s,\textbf{\textit{Q}}}$ for $1s^{-/+}$ trions strongly resemble each other due to comparable effective masses of the conduction and valence bands \cite{kormanyos2015k}.
	A comparison to the radial parts of the $1s$ exciton wave function $\varphi^R\,^{\{K,\uparrow\}}_{1s,\textbf{\textit{Q}}}$, plotted as a black solid line in Fig.~\ref{Bild:wellenfunktionen}(b), shows that the $1s^{-/+}$ trions are more confined in reciprocal space.
	The wave functions $\psi^R\,^{\{K,\uparrow\},\{K',\downarrow\}}_{x\text{-}e/h,-,2s^{-/+},2s,\textbf{\textit{Q}}}$ for $2s^{-/+}$ trions are plotted as blue solid and red dashed lines in Fig.~\ref{Bild:wellenfunktionen}(c) and again strongly resemble each other due to the similar effective masses of the conduction and valence bands.
	The wave functions $\psi^R\,^{\{K,\uparrow\},\{K',\downarrow\}}_{x\text{-}e/h,-,2s^{-/+},2s,\textbf{\textit{Q}}}$ for $2s^{-/+}$ trions are also only slightly less confined in reciprocal space than the $2s$ exciton wave function depicted as a black line in Fig.~\ref{Bild:wellenfunktionen}(c) with a comparable Bohr radius.
	All used material parameters are given in Appendix~\ref{app:parameters} and the treatment of the angular momentum is explained in Appendix~\ref{subsec:magn-quant}.
	While we develop the theory for the whole ensemble of bound and unbound exciton states, our numerical evaluations are restricted to the energetically lowest $\nu = 1s$, $2s$, and $2p^\pm$ exciton states.
	The coefficients~$\hat{T}\,^{\zeta_1,\zeta_2}_{x\text{-}e/h,\pm,\nu,\textbf{\textit{Q}}}$ are now expanded into the basis of wave functions $\psi^R\,^{\zeta_1,\zeta_2}_{x\text{-}e/h,\pm,\mu,\nu,\textbf{\textit{Q}}}$ and new expansion coefficients $T\,^{\zeta_1,\zeta_2}_{x\text{-}e/h,\pm,\mu}$:
	\begin{equation}
	\hat{T}\,^{\zeta_1,\zeta_2}_{x\text{-}e/h,\pm,\nu,\textbf{\textit{Q}}} = \sum_{\mu} \psi^R\,^{\zeta_1,\zeta_2}_{x\text{-}e/h,\pm,\mu,\nu,\textbf{\textit{Q}}} \ T\,^{\zeta_1,\zeta_2}_{x\text{-}e/h,\pm,\mu} . \label{eq:expansion}
	\end{equation}
	The quantum number $\mu$ comprises bound states like $1s^{-/+}$ and $2s^{-/+}$ trions as well as the unbound states forming the exciton-electron/hole continuum.
	As a result, the electron- and hole-density-assisted transitions can be expressed as:
	\begin{widetext}\vspace{-2.7mm}
		\begin{eqnarray}
		& & {\big\langle} c^\dag_{\zeta_1,\textbf{\textit{k}}_1+\alpha_{x\text{-}e}\textbf{\textit{k}}_2+\alpha_{x}\textbf{\textit{Q}}} v^{\phantom{\dagger}}_{\zeta_1,\textbf{\textit{k}}_1-\beta_{x\text{-}e}\textbf{\textit{k}}_2-\beta_{x}\textbf{\textit{Q}}} c^\dag_{\zeta_2,\alpha_{x\text{-}e}\textbf{\textit{k}}_2-\textbf{\textit{Q}}} c^{\phantom{\dagger}}_{\zeta_2,\textbf{\textit{k}}_2} {\big\rangle}^{c} \notag \\
		& & = \sum_{\pm,\mu,\nu} \left( \varphi^R\,^{\zeta_1}_{\nu,\textbf{\textit{k}}_1} \ \psi^R\,^{\zeta_1,\zeta_2}_{x\text{-}e,\pm,\mu,\nu,\textbf{\textit{Q}}} 
		\mp \varphi^R\,^{\zeta_1}_{\nu,\alpha_{x}\textbf{\textit{k}}_1+(\alpha_{x}^2-1)\textbf{\textit{Q}}} \ \psi^R\,^{\zeta_1,\zeta_2}_{x\text{-}e,\pm,\mu,\nu,-\textbf{\textit{k}}_1-\alpha_{x}\textbf{\textit{Q}}} \right) f\,^{\zeta_2}_{e,\textbf{\textit{k}}_2} \ T\,^{\zeta_1,\zeta_2}_{x\text{-}e,\pm,\mu} ,  \label{eq:electron-screened} \\		
		& & {\big\langle} c^\dag_{\zeta_1,\textbf{\textit{k}}_1-\alpha_{x\text{-}h}\textbf{\textit{k}}_2+\alpha_{x}\textbf{\textit{Q}}} v^{\phantom{\dagger}}_{\zeta_1,\textbf{\textit{k}}_1+\beta_{x\text{-}h}\textbf{\textit{k}}_2-\beta_{x}\textbf{\textit{Q}}} v^{\phantom{\dagger}}_{\zeta_2,\beta_{x\text{-}h}\textbf{\textit{k}}_2+\textbf{\textit{Q}}} v^\dag_{\zeta_2,\textbf{\textit{k}}_2} {\big\rangle}^{c} \notag \\
		& & = \sum_{\pm,\mu,\nu} \left(\varphi^R\,^{\zeta_1}_{\nu,\textbf{\textit{k}}_1} \ \psi^R\,^{\zeta_1,\zeta_2}_{x\text{-}h,\pm,\mu,\nu,\textbf{\textit{Q}}} \mp \varphi^R\,^{\zeta_1}_{\nu,\beta_{x}\textbf{\textit{k}}_1+(1-\beta_{x}^2)\textbf{\textit{Q}}} \ \psi^R\,^{\zeta_1,\zeta_2}_{x\text{-}h,\pm,\mu,\nu,\textbf{\textit{k}}_1-\beta_{x}\textbf{\textit{Q}}}\right) f\,^{\zeta_2}_{h,\textbf{\textit{k}}_2} \ T\,^{\zeta_1,\zeta_2}_{x\text{-}h,\pm,\mu} . \label{eq:hole-screened} 
		\end{eqnarray}
\vspace{-4.7mm}\end{widetext}
	
	\begin{figure}[b]
		\centering
		\includegraphics[width=1\columnwidth]{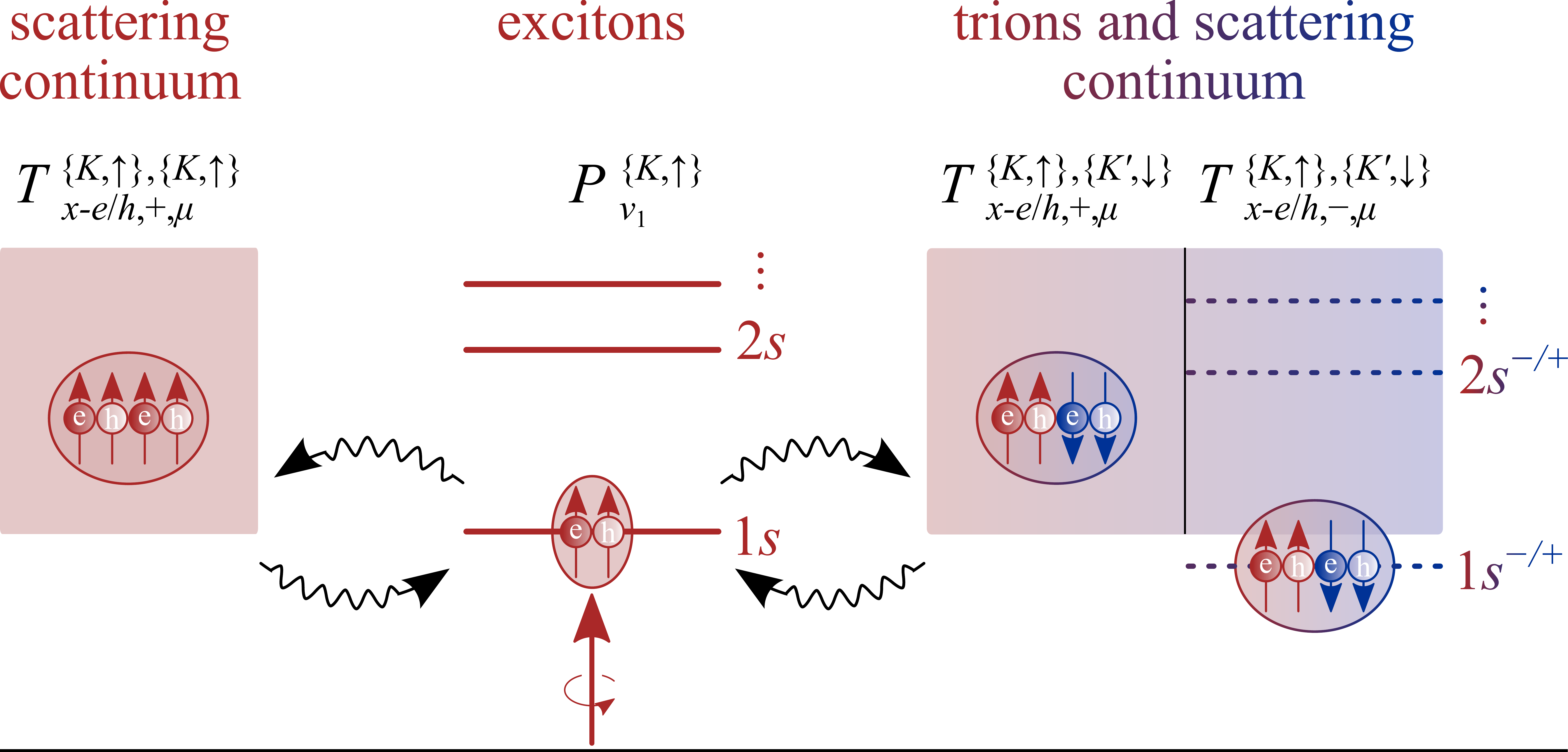}
		\caption{
			Exciton transitions $P\,^{\{K,\uparrow\}}_{\nu_1}$ in the $K$ valley are optically created by an incoming light field illustrated by the red arrow.
			In doped semiconductors, the optically excited exciton transitions $P\,^{\{K,\uparrow\}}_{\nu_1}$ couple via Coulomb interactions to intravalley $T\,^{\{K,\uparrow\},\{K,\uparrow\}}_{x\text{-}e/h,+,\mu\neq t}$ and intervalley $T\,^{\{K,\uparrow\},\{K',\downarrow\}}_{x\text{-}e/h,\pm,\mu\neq t}$  scattering continuua ($\mu \neq t$) as well as intervalley trions $T\,^{\{K,\uparrow\},\{K',\downarrow\}}_{x\text{-}e/h,-,\mu=t}$.
		}
		\label{Bild:schema}
	\end{figure}

\section{Excitonic Equations of Motion} \label{sec:factorized-eqs-of-motion}

	The set of coupled Heisenberg equations of motion characterizing doped atomically thin semiconductors are derived in the following.
	The dynamics of exciton transitions $P\,^{\zeta_1}_{\nu_1}$ is described in Sec.~\ref{subsection:exciton-transitions}.
	In case of doped semiconductors, the exciton transitions $P\,^{\zeta_1}_{\nu_1}$ couple to trion transitions and exciton-electron or exciton-hole continuum transitions $T\,^{\zeta_1,\zeta_2}_{x\text{-}e/h,\pm,\mu}$ as explained in Sec.~\ref{subsection:neg-trion}.
	The interaction is schematically illustrated in Fig.~\ref{Bild:schema}.
	For an in-depth derivation of the set of coupled excitonic Bloch equations see Appendix~\ref{App:derivation}.

\subsection{Exciton Transitions} \label{subsection:exciton-transitions}

	The equation of motion for exciton transitions~$P\,^{\zeta_1}_{\nu_1}$, depicted as red solid lines in Fig.~\ref{Bild:schema}, reads:
	\begin{eqnarray}
	& & \left(\partial_t + \gamma\,^{\zeta_1}_{x} - \frac{i}{\hbar} \epsilon\,^{\zeta_1}_{x,\nu_1} \right) P\,^{\zeta_1}_{\nu_1} \notag \\
	& & = -\frac{i}{\mathcal{A}} \sum_{\textbf{\textit{k}}_1} \Omega\,^{\zeta_1,\sigma_j}_{\nu_1,\textbf{\textit{k}}_1}
	\left(	1 - f\,^{\zeta_1}_{e/h,\textbf{\textit{k}}_1}\right) \notag \\
	& & \hspace{3.8mm} + \frac{i}{\hbar\mathcal{A}} \sum_{\nu_2,\textbf{\textit{k}}_1} \hat{W}\,^{\zeta_1}_{H\text{-}F,\nu_2,\nu_1,\textbf{\textit{k}}_1} \ f\,^{\zeta_1}_{e/h,\textbf{\textit{k}}_1} \ P\,^{\zeta_1}_{\nu_2} \notag \\
	& & \hspace{3.8mm} + \frac{i}{\hbar\mathcal{A}} \sum_{\substack{\zeta_2,\pm \\ \nu_2,\textbf{\textit{k}}_1,\textbf{\textit{k}}_2}} \hat{W}\,^{\zeta_1}_{x\text{-}e/h,\pm,\nu_2,\nu_1,\textbf{\textit{k}}_2,\textbf{\textit{k}}_1} \ f\,^{\zeta_2}_{e/h,\textbf{\textit{k}}_1} \notag \\
	& & \hspace{23.4mm} \times \sum_{\mu} \psi^R\,^{\zeta_1,\zeta_2}_{x\text{-}e/h,\pm,\mu,\nu_2,\textbf{\textit{k}}_2} \ T\,^{\zeta_1,\zeta_2}_{x\text{-}e/h,\pm,\mu} . \arxivcustom
	\label{eq:chi-3-pol}
	\end{eqnarray}
	The left-hand side of Eq.~\eqref{eq:chi-3-pol} describes free excitonic oscillations with the exciton energy $\epsilon\,^{\zeta_1}_{x,\nu_1}$ which are damped by the phonon-mediated dephasing $\gamma\,^{\zeta_1}_{x}$ \cite{selig2016excitonic,christiansen2017phonon,lengers2020theory}.
	An additional radiative dephasing is determined by the self-consistent treatment of the coupled Maxwell's and excitonic Bloch equations \cite{knorr1996theory,jahnke1997linear,katsch2020optical}.
	The first term on the right-hand side of Eq.~\eqref{eq:chi-3-pol} is the optical source term due to an external light field propagating perpendicular to the atomically thin semiconductor.
	The light matter interaction term also includes Pauli blocking proportional to the residual electron or hole doping density $f\,^{\zeta_1}_{e/h,\textbf{\textit{k}}}$.
	The excitonic Rabi frequency~$\sum_{\textbf{\textit{k}}} \Omega\,^{\zeta_1,\sigma_j}_{\nu_1,\textbf{\textit{k}}}$ is defined in Eq.~\eqref{eq:Rabi-freq} and $\mathcal{A}$ denotes the normalization area.
	The second term on the right-hand side of Eq.~\eqref{eq:chi-3-pol} represents a Coulomb-induced exciton energy renormalization which increases the exciton resonance energy depending on the electron or hole doping density $f\,^{\zeta_1}_{e/h,\textbf{\textit{k}}}$.
	The associated Coulomb matrix element $\hat{W}\,^{\zeta_1}_{H\text{-}F,\nu_2,\nu_1,\textbf{\textit{k}}}$ is defined in Eq.~\eqref{eq:mat-el-hf}.
	The third contribution to the right-hand side of Eq.~\eqref{eq:chi-3-pol} describes negatively/positively charged trion transitions ($\mu=t$) and exciton-electron/hole continuum transitions ($\mu\neq t$) $T\,^{\zeta_1,\zeta_2}_{x\text{-}e/h,\pm,\mu}$ as sources for optically generated exciton transitions $P\,^{\zeta_1}_{\nu_1}$.
	Trions and the scattering continua are illustrated as dashed lines and shaded areas in Fig.~\ref{Bild:schema}, respectively.
	The involved Coulomb matrix $\hat{W}\,^{\zeta_1}_{x\text{-}e/h,\pm,\nu_1,\nu_2,\textbf{\textit{k}}_1,\textbf{\textit{k}}_2}$ is defined by Eqs.~\eqref{eq:mat-el-x-e} and \eqref{eq:mat-el-x-h}.
	Note that the description is restricted to electron or hole doping densities and does not characterize the dynamics of optically excited densities which simultaneously occur \cite{steinhoff2014influence,steinhoff2016nonequilibrium,meckbach2018giant,erben2018excitation}.
	Therefore, a finite electron doping density $N_{e}$ implies a vanishing hole doping density $N_{h}$, where only exciton-electron scattering contributes and exciton-hole interactions are zero.
	The opposite holds true for non-zero hole doping densities $N_{h}$.

\subsection{Trion Transitions and Exciton-Electron/Hole Continuum Transitions} \label{subsection:neg-trion}

	The equation of motion for trion transitions ($\mu=t$) and exciton-electron/hole continuum transitions ($\mu\neq t$) $T\,^{\zeta_1,\zeta_2}_{x\text{-}e/h,\pm,\mu}$ reads:
	\begin{eqnarray}
	&& \left[ \partial_t + \gamma\,^{\zeta_1}_{x\text{-}e/h} - \frac{i}{\hbar} \big(\epsilon\,^{\zeta_1,\zeta_2}_{x\text{-}e/h,\pm,\mu} + \Delta\,^{\zeta_2}_{e/h}\big) \right] T\,^{\zeta_1,\zeta_2}_{x\text{-}e/h,\pm,\mu} \notag \\
	& & = \frac{i \left(1\pm\delta_{\zeta_1,\zeta_2}\right)}{2\hbar \mathcal{A} \sum_{\textbf{\textit{k}}_1} f\,^{\zeta_2}_{e/h,\textbf{\textit{k}}_1}} \sum_{\nu_1,\textbf{\textit{k}}_2} \psi^L\,^{\zeta_1,\zeta_2}_{x\text{-}e/h,\pm,\mu,\nu_1,\textbf{\textit{k}}_2} \notag \\
	& & \hspace{3.8mm} \times \sum_{\nu_2,\textbf{\textit{k}}_3} \big(S\,^{\zeta_1}_{x\text{-}e/h,\pm}\big)^{-1}_{\nu_1,\nu_2,\textbf{\textit{k}}_2,\textbf{\textit{k}}_3} \notag \\
	& & \hspace{3.8mm} \times \sum_{\nu_3,\textbf{\textit{k}}_4} \big(\hat{W}\,^{\zeta_1,\zeta_2}_{x\text{-}e/h,\pm,\nu_2,\nu_3,\textbf{\textit{k}}_3,\textbf{\textit{k}}_4}\big)^* \ f\,^{\zeta_2}_{e/h,\textbf{\textit{k}}_4} \ P\,^{\zeta_1}_{\nu_3} . \label{eq:neg-trion-dyn}
	\end{eqnarray}
	%
	The left-hand side of Eq.~\eqref{eq:neg-trion-dyn} represents oscillations with the energy $\epsilon\,^{\zeta_1,\zeta_2}_{x\text{-}e/h,\pm,\mu}$ obtained by a diagonalization of the exciton-electron/hole Coulomb interaction in Eq.~\eqref{eq:Bi-X} which is renormalized by the shift $\Delta\,^{\zeta_2}_{e/h}$ defined in Eqs.~\eqref{eq:e-renorm} and \eqref{eq:h-renorm}.
	The oscillations are damped by a phonon-mediated dephasing $\gamma\,^{\zeta_1}_{x\text{-}e/h}$.
	The contributions on the right-hand side of Eq.~\eqref{eq:neg-trion-dyn} characterize Coulomb-mediated source terms of the trion transitions ($\mu=t$) and exciton-electron/hole continuum transitions ($\mu\neq t$) $T\,^{\zeta_1,\zeta_2}_{x\text{-}e/h,\pm,\mu}$ due to exciton transitions $P\,^{\zeta_1}_{\nu_3}$ and an electron or hole doping density $f\,^{\zeta_2}_{e/h,\textbf{\textit{k}}_4}$.
	Filling factors $(1–f\,^{\zeta}_{e/h,\textbf{\textit{k}}})$ contributing to Eq.~\eqref{eq:neg-trion-dyn} were neglected due to a systematic truncation to linear doping densities valid for doping densities Ne/h and trion Bohr radii $a_t$ satisfying $N_{e/h} (a_t)^2 \ll 1$ \cite{esser2001theory}.
	This condition is fulfilled in the range of low doping densities, where filling factors in Eq.~\eqref{eq:neg-trion-dyn} effectively enter the exciton dynamics, Eq.~\eqref{eq:chi-3-pol}, nonlinear in the doping density $f\,^{\zeta}_{e/h,\textbf{\textit{k}}}$.
	The nonlinear doping dependence results from the coupling of exciton transitions to trions and exciton-electron/hole continua, described by the last contribution to Eq.~\eqref{eq:chi-3-pol}, which also includes the doping density.
	\mbox{However}, for larger doping densities $N_{e/h}$ not only six-particle correlations resulting in additional filling factors \cite{rana2020many} but also dynamical screening \cite{van2017marrying,van2019probing} become of importance.

\section{Doping-Dependent Absorption} \label{sec:results}
		
	\begin{figure*}
		\centering
		\includegraphics[width=\textwidth]{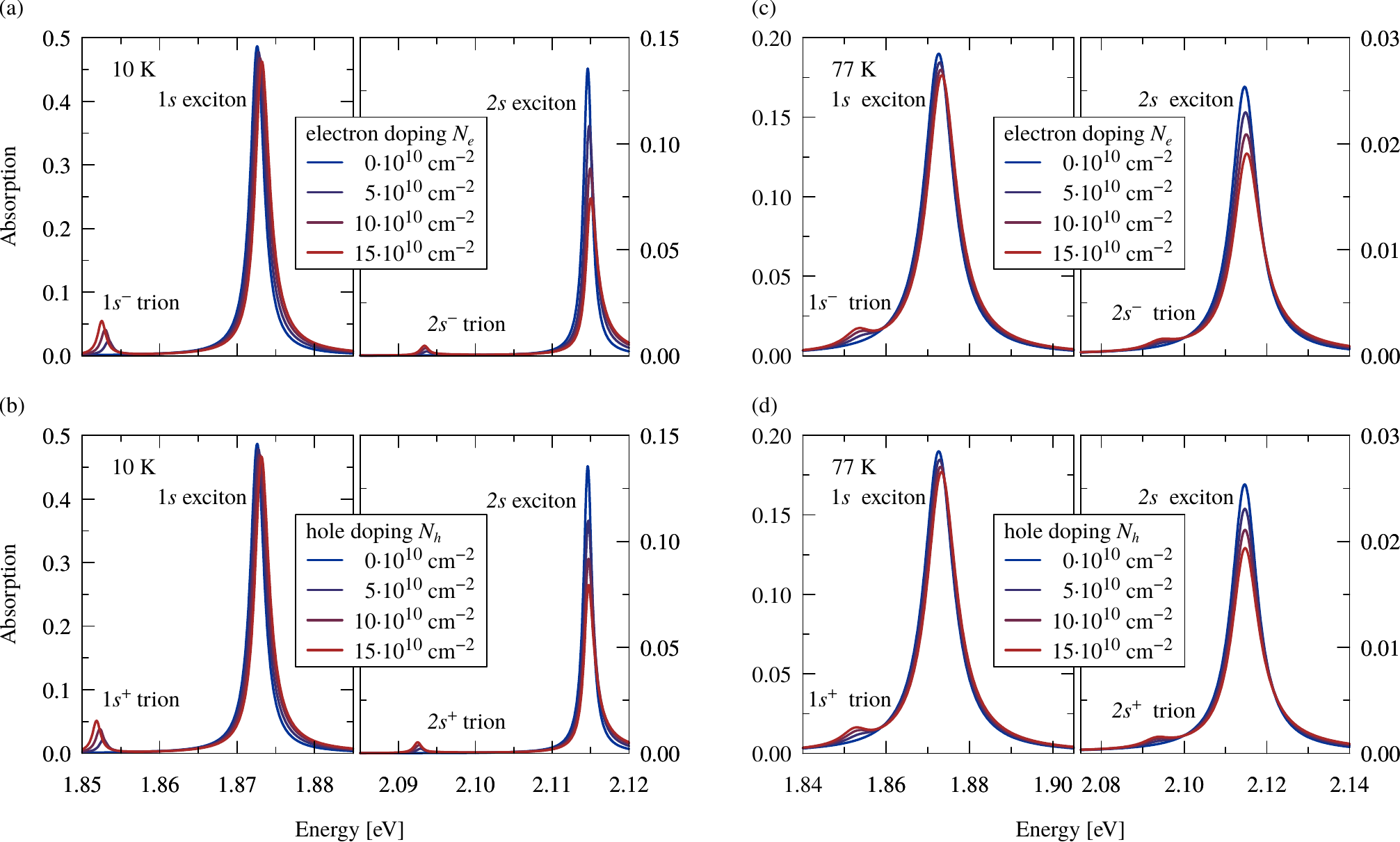}
		\caption{
			Absorption spectra for monolayer MoSe\textsubscript{2} encapsulated in hexagonal BN at (a),(b)~10~K and (c),(d)~77~K.
			The spectra are shown for different (a),(c)~electron doping densities $N_e$ and (b),(d)~hole doping densities $N_h$.
			The $1s$ and $2s$ exciton resonances as well as negatively/positively charged $1s^{-/+}$ and $2s^{-/+}$ trions are indicated.			
		}
		\label{Bild:Refl}
	\end{figure*}

\begin{figure}
		\centering
		\includegraphics[width=1\columnwidth]{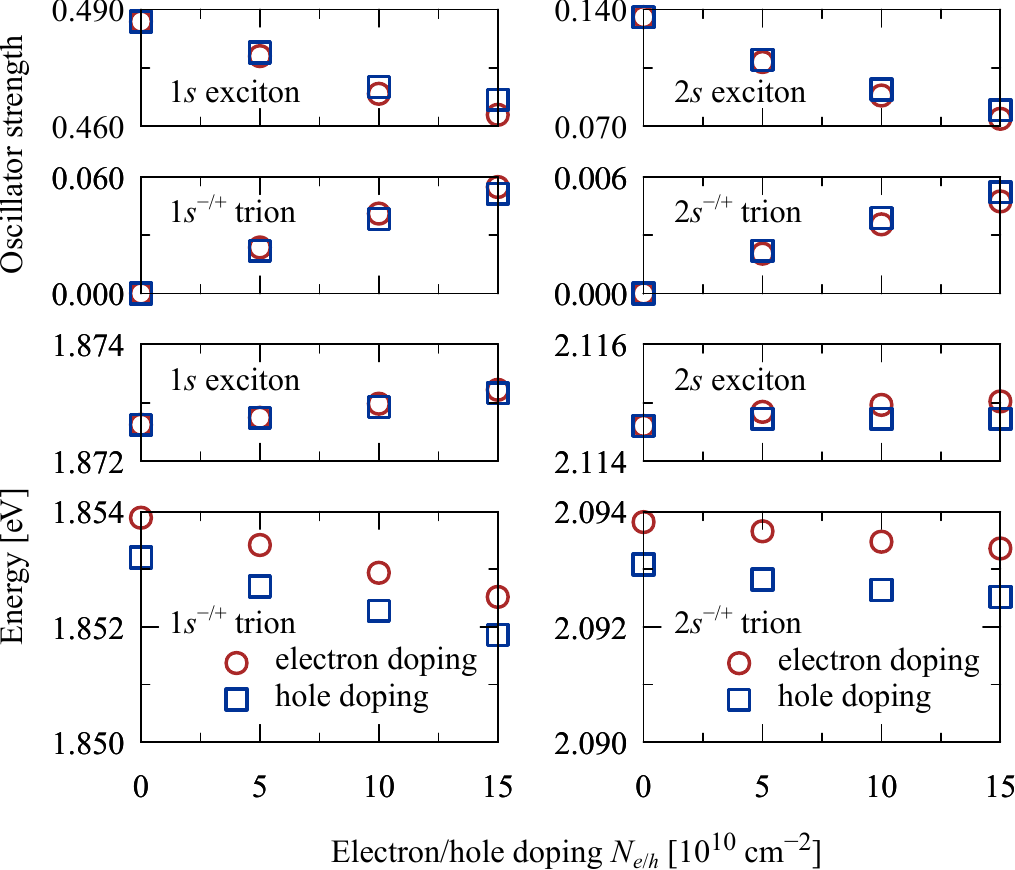}
		\caption{
			Oscillator strengths and resonance energies of $1s$ and $2s$ exciton resonances as well as negatively/positively charged $1s^{-/+}$ and $2s^{-/+}$ trions extracted from the absorption spectra for monolayer MoSe\textsubscript{2} encapsulated in hexagonal BN at 10~K, cf. Figs.~\ref{Bild:Refl}(a),(b).
		}
		\label{Bild:extr}
	\end{figure}

	In the following we discuss the doping-induced changes of the absorption spectra near the energetically lowest $1s$ and $2s$ exciton resonances.
	The absorption spectra at different doping densities are obtained by self-consistently solving the Maxwell's \cite{knorr1996theory,jahnke1997linear} and excitonic Bloch equations, Eqs.~\eqref{eq:chi-3-pol} and \eqref{eq:neg-trion-dyn}.
	The set of coupled excitonic Bloch equations can be solved numerically in time domain or analytically in frequency domain as done in Appendix~\ref{app:fourier-transform}.
	The calculated absorption spectra for monolayer MoSe\textsubscript{2} encapsulated in hexagonal BN as a prototypical atomically thin semiconductor are presented in Fig.~\ref{Bild:Refl}.
	The absorption spectra are evaluated at a temperature of 10~K in Figs.~\ref{Bild:Refl}(a) and \ref{Bild:Refl}(b) and 77~K in Figs.~\ref{Bild:Refl}(c) and \ref{Bild:Refl}(d).
	The red curves depict the absorption for an undoped sample with pronounced energetically lowest $1s$ and excited-state $2s$ exciton resonances.
	Note the different scaling of the absorption for $1s$ and $2s$ excitons in Fig.~\ref{Bild:Refl}.
	The $1s$ exciton linewidth, presented as full width at half maximum, includes a radiative part of approximately 1~meV as well as a phonon-mediated part of about 1~meV at 10~K and 7~meV at 77~K \cite{selig2016excitonic}.
	In contrast, the $2s$ exciton linewidth is mostly dominated by the phonon-mediated part, since the radiative dephasing of $2s$ excitons is much smaller \cite{brem2019intrinsic}.
	The absorption spectra for increasing electron doping densities $N_e$ are plotted in Figs.~\ref{Bild:Refl}(a) and \ref{Bild:Refl}(c) at 10~K and 77~K, respectively.
	With a growing electron doping density $N_e$, the $1s$ and $2s$ exciton oscillator strengths decrease and the resonance energies are slightly shifted toward higher energies compared to the undoped case.
	The oscillator strengths and resonance energies extracted from Fig.~\ref{Bild:Refl}(a) are plotted as red circles in Fig.~\ref{Bild:extr}.
	The reduced oscillator strengths stem from Pauli blocking and a Coulomb-mediated redistribution of the oscillator strength.
	Pauli blocking is described by the first term on the right-hand side of Eq.~\eqref{eq:chi-3-pol}, whereas the Coulomb-mediated redistribution originates from the second and third terms on the right-hand side of Eq.~\eqref{eq:chi-3-pol}.
	The Coulomb-mediated redistribution results in asymmetric exciton line shapes \cite{companion2} and the formation of negatively charged $1s^-$ and $2s^-$ trion resonances, which appear approximately 20~meV below the neutral $1s$ and $2s$ exciton resonances.
	Even though the trion linewidths are governed by individual phonon-mediated dephasing rates, we assumed identical linewidths for all trion and exciton-electron/hole continuum states in a first approximation and adjusted the values to the phonon-mediated dephasing of excitons.
	In particular, we assume equal  $1s^{-/+}$ and $2s^{-/+}$ trion linewidths, but want to emphasize a recent study which found increased $2s^{-/+}$ trion linewidths \cite{wagner2020autoionization}.
	The exciton-trion level repulsion contributes to increasing exciton resonance energies and decreasing trion resonance energies plotted as red circles in Fig.~\ref{Bild:extr}.
	The observation of reduced exciton oscillator strengths, the emergence of trion resonances, as well as exciton and trion energy renormalizations are in agreement with recent measurements on different monolayer TMDCs \cite{wagner2020autoionization,xiao2020many,liu2020gate}.
	Finally, the absorption spectra for an increasing hole doping density $N_h$ are plotted in Figs.~\ref{Bild:Refl}(b) and \ref{Bild:Refl}(d) at 10~K and 77~K, respectively.
	Compared to an undoped sample, the oscillator strengths of neutral $1s$ and $2s$ excitons reduce and their resonance energies shift toward higher energies with rising doping densities.
	The oscillator strengths and resonance energies extracted from Fig.~\ref{Bild:Refl}(b) are plotted as blue squares in Fig.~\ref{Bild:extr}.
	Moreover, positively charged $1s^+$ and $2s^+$ trion resonances appear energetically below the neutral $1s$ and $2s$ excitons, respectively.
	All in all, the absorption is qualitatively equivalent to the previously discussed case of electron doping.
	Again, our observations closely align with recent gate-dependent measurements 
	\cite{goldstein2020ground,wagner2020autoionization,xiao2020many,liu2020gate}.

\section{Conclusion and Outlook} \label{sec:conclusion}

	We introduced a microscopic theory to describe the doping-dependent excitonic properties of atomically thin semiconductors dominated by Coulomb scattering of excitons with residual doping densities.
	Our formalism gives access to the binding energies of negatively and positively charged ground- and excited-state trions as well as the corresponding exciton-electron and exciton-hole scattering continua.
	Thus, our theory provides a basis to explore the fundamental properties of ground and excited-state excitons and trions in semiconductors under the influence of residual doping densities.
	As a first application, we studied the absorption spectra for doped monolayer MoSe\textsubscript{2}.
	An extension of our theory to photoluminescence, which is determined by the photon flux of the emitted light, would require a fully quantized light-matter interaction \cite{kira2006many}.
	Although the exciton and trion energies and wave functions remain the same as inputs to the photoluminescence, the dynamics needs to be extended to account for scattering-induced incoherent exciton and trion occupations in addition to the coherences that govern the linear absorption.
	Moreover, phonon-assisted processes must often be considered when describing photoluminescence \cite{selig2017dark}.
	The theoretical formalism can be adapted to other inorganic or organic semiconductors which due to strong Coulomb interactions exhibit tightly bound excitons.
	Possible candidates are not only atomically thin semiconductors, like van der Waals bound homobilayers \cite{horng2018observation,deilmann2018interlayer2,niehues2019interlayer,paradisanos2020controlling} and heterobilayers of TMDCs \cite{rivera2015observation,miller2017long,deilmann2018interlayer,tran2019evidence,alexeev2019resonantly,brem2020hybridized,brem2020tunable}, but also quantum wires or bulk materials.
	Concurrently, the simplicity of our theory allows further adaptations, for instance, to study the trion dynamics in optical wave mixing experiments \cite{singh2016trion,hao2016coherent,Venanzi2021tera,kwong2021effect,Rana2021steady} or photoluminescence \cite{wang2015polarization,plechinger2016trion,godde2016exciton,robert2016exciton} influenced by phonon-assisted relaxation phenomena \cite{selig2019ultrafast,christiansen2019theory}.
	Further perspectives could be to theoretically investigate the influence of doping on spatiotemporal dynamic effects not only in monolayer TMDCs \cite{kato2016transport,yuan2017exciton,kulig2018exciton,rosati2018spatial,perea2019exciton,rosati2019effective,zipfel2020exciton,lengers2020phonon} but also in other atomically thin semiconductors like hybrid perovskites \cite{deng2020long,seitz2020exciton,ziegler2020fast}.

	\begin{acknowledgements}
		We thank Dominik Christiansen and Malte Selig (TU Berlin) for many stimulating discussions. We gratefully acknowledge support from the Deutsche Forschungsgemeinschaft through Project No. 420760124 (\mbox{KN 427/11-1}).
	\end{acknowledgements}

\appendix
\section{Screened Coulomb Potential} \label{app:screen}	

	The two-dimensional Coulomb potential $V_\textbf{\textit{k}}$ including the elementary charge $e_0$, the vacuum permittivity $\varepsilon_0$, and the two-dimensional normalization area $\mathcal{A}$ is described by:
	\begin{equation}
	V_\textbf{\textit{k}} = \frac{e_0^2}{2\varepsilon_0\mathcal{A}|\textbf{\textit{k}}|} . \label{eq:Coulomb-pot}
	\end{equation}
	The bare Coulomb potential $V_\textbf{\textit{k}}$ is screened by $\varepsilon_\textbf{\textit{k}}$ which takes account of dielectric screening from an encapsulating material characterized by the constant $\varepsilon_e$ \cite{florian2018dielectric,steinhoff2020dynamical}:
	\begin{equation}
	\varepsilon_\textbf{\textit{k}} = 
	\varepsilon_{2d}\frac{1-\tilde{\varepsilon}_1 e^{-(h_{2d}+2h_{g}) |\textbf{\textit{k}}|}-\tilde{\varepsilon}_2 e^{-h_{2d}|\textbf{\textit{k}}|}+\tilde{\varepsilon}_1\tilde{\varepsilon}_2e^{-2h_{g} |\textbf{\textit{k}}|}}{1+\tilde{\varepsilon}_1 e^{-(h_{2d}+2h_{g}) |\textbf{\textit{k}}|}+\tilde{\varepsilon}_2 e^{-h_{2d}|\textbf{\textit{k}}|}+\tilde{\varepsilon}_1\tilde{\varepsilon}_2e^{-2h_{g} |\textbf{\textit{k}}|}} ,
	\label{eq:screening}
	\end{equation}
	with the dielectric constant $\varepsilon_{2d}$ and the height $h_{2d}$ of the atomically thin semiconductor as well as the abbreviations $\tilde{\varepsilon}_1 = (1-\varepsilon_e)/(1+\varepsilon_e)$ and $\tilde{\varepsilon}_2 = (\varepsilon_{2d}-1)/(\varepsilon_{2d}+1)$.
	The dielectric screening described by Eq.~\eqref{eq:screening} includes naturally occurring air gaps \cite{rooney2017observing} between the atomically thin semiconductor and its dielectric environment characterized by the dielectric constant $\varepsilon_e$.
	These interlayer gaps of width $h_{g}$ were assumed to be identical on both side of the atomically thin semiconductor.
	%

\section{Expansion of Hole-Density-Assisted Transitions} \label{subsec:pos-obs}

	The treatment of hole-density-assisted transitions ${\big\langle} c^\dag_{\zeta_1,\textbf{\textit{k}}_1+\textbf{\textit{Q}}} v^{\phantom{\dagger}}_{\zeta_1,\textbf{\textit{k}}_1} v^{\phantom{\dagger}}_{\zeta_2,\textbf{\textit{k}}_2+\textbf{\textit{Q}}} v^\dag_{\zeta_2,\textbf{\textit{k}}_2} {\big\rangle}^{c}$ is similar to electron-density-assisted transitions ${\big\langle} c^\dag_{\zeta_1,\textbf{\textit{k}}_1+\textbf{\textit{Q}}} v^{\phantom{\dagger}}_{\zeta_1,\textbf{\textit{k}}_1} c^\dag_{\zeta_2,\textbf{\textit{k}}_2-\textbf{\textit{Q}}} c^{\phantom{\dagger}}_{\zeta_2,\textbf{\textit{k}}_2} {\big\rangle}^{c}$ discussed in Sec.~\ref{subsec:neg-obs}.

	The antisymmetry with respect to the exchange of the two valence band annihilation operators suggests the following expansion:	
	\onecolumngrid
	\noindent 
	\begin{eqnarray}
	& & {\big\langle} c^\dag_{\zeta_1,\textbf{\textit{k}}_1-\alpha_{x\text{-}h}\textbf{\textit{k}}_2+\alpha_{x}\textbf{\textit{Q}}} v^{\phantom{\dagger}}_{\zeta_2,\textbf{\textit{k}}_1+\beta_{x\text{-}h}\textbf{\textit{k}}_2-\beta_{x}\textbf{\textit{Q}}} v^{\phantom{\dagger}}_{\zeta_3,\beta_{x\text{-}h}\textbf{\textit{k}}_2+\textbf{\textit{Q}}} v^\dag_{\zeta_4,\textbf{\textit{k}}_2} {\big\rangle}^{c} \notag \\
	& & = \sum_{\nu} \Big( \varphi^R\,^{\zeta_1}_{\nu,\textbf{\textit{k}}_1} \ \mathcal{T}\,^{\zeta_1,\zeta_2,\zeta_3,\zeta_4}_{x\text{-}h,\nu,\textbf{\textit{Q}}} 
	- \varphi^R\,^{\zeta_1}_{\nu,\beta_{x}\textbf{\textit{k}}_1+[1-\beta_{x}^2]\textbf{\textit{Q}}} \ \mathcal{T}\,^{\zeta_1,\zeta_3,\zeta_2,\zeta_4}_{x\text{-}h,\nu,\textbf{\textit{k}}_1-\beta_{x}\textbf{\textit{Q}}} \Big) f\,^{\zeta_4}_{h,\textbf{\textit{k}}_2} .
	\end{eqnarray}
	Again, \mbox{``$+$''} symmetric and \mbox{``$-$''} antisymmetric linear combinations are treated:
	\begin{eqnarray}
		& & \frac{1}{2}\left({\big\langle} c^\dag_{\zeta_1,\textbf{\textit{k}}_1-\alpha_{x\text{-}h}\textbf{\textit{k}}_2+\alpha_{x}\textbf{\textit{Q}}} v^{\phantom{\dagger}}_{\zeta_1,\textbf{\textit{k}}_1+\beta_{x\text{-}h}\textbf{\textit{k}}_2-\beta_{x}\textbf{\textit{Q}}} v^{\phantom{\dagger}}_{\zeta_2,\beta_{x\text{-}h}\textbf{\textit{k}}_2+\textbf{\textit{Q}}} v^\dag_{\zeta_2,\textbf{\textit{k}}_2} {\big\rangle}^{c}
		\pm {\big\langle} c^\dag_{\zeta_1,\textbf{\textit{k}}_1-\alpha_{x\text{-}h}\textbf{\textit{k}}_2+\alpha_{x}\textbf{\textit{Q}}} v^{\phantom{\dagger}}_{\zeta_2,\textbf{\textit{k}}_1+\beta_{x\text{-}h}\textbf{\textit{k}}_2-\beta_{x}\textbf{\textit{Q}}} v^{\phantom{\dagger}}_{\zeta_1,\beta_{x\text{-}h}\textbf{\textit{k}}_2+\textbf{\textit{Q}}} v^\dag_{\zeta_2,\textbf{\textit{k}}_2} {\big\rangle}^{c}\right) \notag \\
		& & = \sum_{\nu} \left(\varphi^R\,^{\zeta_1}_{\nu,\textbf{\textit{k}}_1} \ \hat{T}\,^{\zeta_1,\zeta_2}_{x\text{-}h,\pm,\nu,\textbf{\textit{Q}}}
		\mp \varphi^R\,^{\zeta_1}_{\nu,\beta_{x}\textbf{\textit{k}}_1+[1-\beta_{x}^2]\textbf{\textit{Q}}} \ \hat{T}\,^{\zeta_1,\zeta_2}_{x\text{-}h,\pm,\nu,\textbf{\textit{k}}_1-\beta_{x}\textbf{\textit{Q}}}\right)f\,^{\zeta_2}_{h,\textbf{\textit{k}}_2}  ,
	\end{eqnarray}	
	with the expansion coefficients $\hat{T}\,^{\zeta_1,\zeta_2}_{x\text{-}h,\nu,\textbf{\textit{Q}}}$ defined by:
	$\hat{T}\,^{\zeta_1,\zeta_2}_{x\text{-}h,\nu,\textbf{\textit{Q}}} = \frac{1}{2} \big(\mathcal{T}\,^{\zeta_1,\zeta_1,\zeta_2,\zeta_2}_{x\text{-}h,\nu,\textbf{\textit{Q}}} \pm \mathcal{T}\,^{\zeta_1,\zeta_2,\zeta_1,\zeta_2}_{x\text{-}h,\nu,\textbf{\textit{Q}}}\big)$.

\section{Matrix Elements} \label{app:parameters}

	In the following, the used matrix elements are defined:
	The matrices $S\,^{\zeta_1}_{x\text{-}e/h,\pm,\nu_1,\nu_2,\textbf{\textit{Q}}_1,\textbf{\textit{Q}}_2}$ and the Coulomb interaction kernels $W\,^{\zeta_1}_{x\text{-}e/h,\pm,\nu_1,\nu_2,\textbf{\textit{Q}}_1,\textbf{\textit{Q}}_2}$ which determine Eq.~\eqref{eq:Bi-X} are defined by:
	\begin{eqnarray}
	S\,^{\zeta_1}_{x\text{-}e,\pm,\nu_1,\nu_2,\textbf{\textit{Q}}_1,\textbf{\textit{Q}}_2} & = & \delta_{\nu_1,\nu_2} \ \delta_{\textbf{\textit{Q}}_1,\textbf{\textit{Q}}_2} \mp \frac{1}{\mathcal{A}} \ \varphi^L\,^{\zeta_1}_{\nu_1,-\alpha_{x}\textbf{\textit{Q}}_1-\textbf{\textit{Q}}_2} \ \varphi^R\,^{\zeta_1}_{\nu_2,-\textbf{\textit{Q}}_1-\alpha_{x}\textbf{\textit{Q}}_2}
	, \label{eq:S-Matrix} \\
	S\,^{\zeta_1}_{x\text{-}h,\pm,\nu_1,\nu_2,\textbf{\textit{Q}}_1,\textbf{\textit{Q}}_2} & = & \delta_{\nu_1,\nu_2} \ \delta_{\textbf{\textit{Q}}_1,\textbf{\textit{Q}}_2} \mp \frac{1}{\mathcal{A}} \ \varphi^L\,^{\zeta_1}_{\nu_1,\beta_{x}\textbf{\textit{Q}}_1+\textbf{\textit{Q}}_2} \ \varphi^R\,^{\zeta_1}_{\nu_2,\textbf{\textit{Q}}_1+\beta_{x}\textbf{\textit{Q}}_2} , \label{eq:S-Matrix2} \\
	W\,^{\zeta_1}_{x\text{-}e,\pm,\nu_1,\nu_2,\textbf{\textit{Q}}_1,\textbf{\textit{Q}}_2} & = & \frac{1}{\mathcal{A}} \sum_{\textbf{\textit{k}}} \varphi^L\,^{\zeta_1}_{\nu_1,\textbf{\textit{k}}} \left[W_{\textbf{\textit{Q}}_1-\textbf{\textit{Q}}_2} \left(\varphi^R\,^{\zeta_1}_{\nu_2,\textbf{\textit{k}}-\beta_{x}(\textbf{\textit{Q}}_1-\textbf{\textit{Q}}_2)} - \varphi^R\,^{\zeta_1}_{\nu_2,\textbf{\textit{k}}+\alpha_{x}(\textbf{\textit{Q}}_1-\textbf{\textit{Q}}_2)}\right) \right. \notag \\
	& & \hspace{22.0mm} \left. \mp W_{\textbf{\textit{k}}+\alpha_{x}\textbf{\textit{Q}}_1+\textbf{\textit{Q}}_2} \left(\varphi^R\,^{\zeta_1}_{\nu_2,\textbf{\textit{k}}-\beta_{x}(\textbf{\textit{Q}}_1-\textbf{\textit{Q}}_2)} - \varphi^R\,^{\zeta_1}_{\nu_2,-\textbf{\textit{Q}}_1-\alpha_{x}\textbf{\textit{Q}}_2}\right)\right] , 
	\label{eq:Coulomb-M-E} \\
	W\,^{\zeta_1}_{x\text{-}h,\pm,\nu_1,\nu_2,\textbf{\textit{Q}}_1,\textbf{\textit{Q}}_2} & = & \frac{1}{\mathcal{A}} \sum_{\textbf{\textit{k}}} \varphi^L\,^{\zeta_1}_{\nu_1,\textbf{\textit{k}}} \left[W_{\textbf{\textit{Q}}_1-\textbf{\textit{Q}}_2} \left(\varphi^R\,^{\zeta_1}_{\nu_2,\textbf{\textit{k}}+\alpha_{x}(\textbf{\textit{Q}}_1-\textbf{\textit{Q}}_2)} - \varphi^R\,^{\zeta_1}_{\nu_2,\textbf{\textit{k}}-\beta_{x}(\textbf{\textit{Q}}_1-\textbf{\textit{Q}}_2)}\right)
	\right. \notag \\
	& & \hspace{22.0mm} \left. \mp W_{-\textbf{\textit{k}}+\beta_{x}\textbf{\textit{Q}}_1+\textbf{\textit{Q}}_2} \left(\varphi^R\,^{\zeta_1}_{\nu_2,\textbf{\textit{k}}+\alpha_{x}(\textbf{\textit{Q}}_1-\textbf{\textit{Q}}_2)} - \varphi^R\,^{\zeta_1}_{\nu_2,\textbf{\textit{Q}}_1+\beta_{x}\textbf{\textit{Q}}_2}\right)\right] .  \label{eq:Coulomb-M-E-2}
	\end{eqnarray}
	The light-matter interaction strength is characterized by the Rabi frequency $\Omega\,^{\zeta_1,\sigma_j}_{\nu_1,\textbf{\textit{k}}}$:
	\begin{equation}
	\Omega\,^{\zeta_1,\sigma_j}_{\nu_1,\textbf{\textit{k}}}(t) = \frac{1}{\hbar} \ \varphi^L\,^{\zeta_1}_{\nu_1,\textbf{\textit{k}}} \big[ d\,^{\zeta_1,\sigma_j}_{c,v\vphantom{\textbf{\textit{k}}_1}} \ \tilde{E}_T^{\sigma_j}(t) \ e^{-i \omega_0 t} \big]^* . \label{eq:Rabi-freq}
	\end{equation}
	The interband dipole transition element $d\,^{\zeta_1,\sigma_j}_{c,v}$ involves the elementary charge $e_0$, the material parameter $\gamma_{2d}$, and the band gap $\varepsilon\,^{\zeta_1}_g$:
	\begin{equation}
	d\,^{\zeta_1=\{\xi_1,s_1\},\sigma_j}_{c,v} = - i (\delta_{\sigma_j,\sigma_+}\delta_{\xi_1,K} + \delta_{\sigma_j,\sigma_-}\delta_{\xi_1,K'}) \, \frac{\sqrt{2} e_0 \gamma_{2d}}{\varepsilon\,^{\zeta_1}_g} . \label{eq:dipole-ele}
	\end{equation}
	The Kronecker deltas take account of the valley-selective circular dichroism in monolayer TMDCs \cite{yao2008valley,cao2012valley,zeng2012valley,mak2012control,xiao2012coupled}.
	Equation~\eqref{eq:Rabi-freq} additionally includes the envelope of the light field at the monolayer position  $\smash[t]{\big[\tilde{E}_T^{\sigma_j}(t)\big]^*}$ as well as the phase factor $e^{i \omega_0 t}$ determined by the optical frequency $\omega_0$.
	The Hartree--Fock Coulomb matrix element $\hat{W}\,^{\zeta_1}_{H\text{-}F,\nu_1,\nu_2,\textbf{\textit{Q}}} $ is defined by:
	\begin{eqnarray}
	\hat{W}\,^{\zeta_1}_{H\text{-}F,\nu_1,\nu_2,\textbf{\textit{Q}}} & = & \sum_{\textbf{\textit{k}}} W_{\textbf{\textit{k}}-\textbf{\textit{Q}}} \ \varphi^R\,^{\zeta_1}_{\nu_1,\textbf{\textit{k}}} \left(\varphi^L\,^{\zeta_1}_{\nu_2,\textbf{\textit{Q}}} - \varphi^L\,^{\zeta_1}_{\nu_2,\textbf{\textit{k}}}\right) .  \label{eq:mat-el-hf}
	\end{eqnarray}
	The Coulomb matrices $\hat{W}\,^{\zeta_1}_{x\text{-}e/h,\pm,\nu_1,\nu_2,\textbf{\textit{Q}}_1,\textbf{\textit{Q}}_2}$ describing the coupling between excitons and trions as well as exciton-electron/hole continuum states in Eqs.~\eqref{eq:chi-3-pol} and \eqref{eq:neg-trion-dyn} are given by:
	\begin{eqnarray}
	\hat{W}\,^{\zeta_1}_{x\text{-}e,\pm,\nu_1,\nu_2,\textbf{\textit{Q}}_1,\textbf{\textit{Q}}_2} & = & \sum_{\textbf{\textit{k}}} \varphi^R\,^{\zeta_1}_{\nu_1,\textbf{\textit{k}}} \left[W_{\textbf{\textit{Q}}_1+(1-\alpha_{x\text{-}e})\textbf{\textit{Q}}_2} \left(\varphi^L\,^{\zeta_1}_{\nu_2,\textbf{\textit{k}}-\beta_{x}\textbf{\textit{Q}}_1-\beta_{x\text{-}e}\textbf{\textit{Q}}_2} - \varphi^L\,^{\zeta_1}_{\nu_2,\textbf{\textit{k}}+\alpha_{x}\textbf{\textit{Q}}_1+\alpha_{x\text{-}e}\textbf{\textit{Q}}_2}\right) \right.\notag \\
	& & \hspace{18.0mm} \left. \mp W_{\textbf{\textit{k}}+\alpha_{x}\textbf{\textit{Q}}_1-(1-\alpha_{x\text{-}e})\textbf{\textit{Q}}_2} \left(\varphi^L\,^{\zeta_1}_{\nu_2,\textbf{\textit{k}}-\beta_{x}\textbf{\textit{Q}}_1-\beta_{x\text{-}e}\textbf{\textit{Q}}_2} - \varphi^L\,^{\zeta_1}_{\nu_2,-\textbf{\textit{Q}}_1+\alpha_{x\text{-}e}\textbf{\textit{Q}}_2}\right)\right] .  \label{eq:mat-el-x-e} \\
	\hat{W}\,^{\zeta_1}_{x\text{-}h,\pm,\nu_1,\nu_2,\textbf{\textit{Q}}_1,\textbf{\textit{Q}}_2} & = & \sum_{\textbf{\textit{k}}} \varphi^R\,^{\zeta_1}_{\nu_1,\textbf{\textit{k}}} \left[W_{\textbf{\textit{Q}}_1-(1-\beta_{x\text{-}h})\textbf{\textit{Q}}_2} \left(\varphi^L\,^{\zeta_1}_{\nu_2,\textbf{\textit{k}}+\alpha_{x}\textbf{\textit{Q}}_1-\alpha_{x\text{-}h}\textbf{\textit{Q}}_2} - \varphi^L\,^{\zeta_1}_{\nu_2,\textbf{\textit{k}}-\beta_{x}\textbf{\textit{Q}}_1+\beta_{x\text{-}h}\textbf{\textit{Q}}_2}\right) \right. \notag \\
	& & \hspace{18.0mm} \left. \mp W_{-\textbf{\textit{k}}+\beta_{x}\textbf{\textit{Q}}_1+(1-\beta_{x\text{-}h})\textbf{\textit{Q}}_2} \left(\varphi^L\,^{\zeta_1}_{\nu_2,\textbf{\textit{k}}+\alpha_{x}\textbf{\textit{Q}}_1-\alpha_{x\text{-}h}\textbf{\textit{Q}}_2} - \varphi^L\,^{\zeta_1}_{\nu_2,\textbf{\textit{Q}}_1+\beta_{x\text{-}h}\textbf{\textit{Q}}_2}\right)\right] .  \label{eq:mat-el-x-h}
	\end{eqnarray}
	Finally, the electron ``$e$'' and hole ``$h$'' renormalizations $\Delta\,^{\zeta_2}_{e/h}$ appearing in Eq.~\eqref{eq:neg-trion-dyn} read:
	\begin{eqnarray}
	\Delta\,^{\zeta_2}_{e} & = & \frac{1}{\sum_{\textbf{\textit{k}}_1} f\,^{\zeta_2}_{e,\textbf{\textit{k}}_1}} \sum_{\textbf{\textit{k}}_2} \frac{\hbar^2\textbf{\textit{k}}_2^2}{2} \frac{m_{e}+m_{h}}{(2m_{e}+m_{h})m_{e}} f\,^{\zeta_2}_{e,\textbf{\textit{k}}_2} , \label{eq:e-renorm} \\
	\Delta\,^{\zeta_2}_{h} & = & \frac{1}{\sum_{\textbf{\textit{k}}_1} f\,^{\zeta_2}_{h,\textbf{\textit{k}}_1}} \sum_{\textbf{\textit{k}}_2} \frac{\hbar^2\textbf{\textit{k}}_2^2}{2}\frac{m_{e}+m_{h}}{(m_{e}+2m_{h})m_{h}} f\,^{\zeta_2}_{h,\textbf{\textit{k}}_2} . \label{eq:h-renorm}
	\end{eqnarray}	

	All used material parameters for monolayer MoSe\textsubscript{2} are listed in Tab.~\ref{Tab:Parameters-mose2}.
	\begin{table}
		\caption{Material parameters for monolayer MoSe\textsubscript{2}.}
		\centering
		\begin{tabular}{lll}
			\hline
			Thickness & $d_{2d}$ & 0.668~nm  \cite{rasmussen2015computational} \\
			Single particle band gap & $\varepsilon_g$ & 2.18~eV \cite{rasmussen2015computational} \\
			Effective electron mass & $m_{e}$ & 0.49~$m_0$ \cite{kormanyos2015k} \\
			Effective hole mass & $m_{h}$ & 0.59~$m_0$ \cite{kormanyos2015k} \\ 
			In-plane dielectric constant & $\epsilon_{2d}$ & 13.3 \cite{kumar2012tunable} \\
			Material parameter & $\gamma_{2d}$ & 0.22~eV~nm \cite{kormanyos2015k} \\
			Interlayer gap & $h_{g}$ & 0.3~nm \\
			Phonon-mediated dephasing \hspace{10mm} & $\hbar\gamma\,^{\{K,\uparrow\}}_{x}=\hbar\gamma\,^{\{K',\downarrow\}}_{x}$  \hspace{10mm} & 0.6/3.8~meV at 10/77~K \cite{blueshift} \\
			\hline
		\end{tabular}
		\label{Tab:Parameters-mose2}
	\end{table}

\section{Derivation of Excitonic Bloch Equations} \label{App:derivation}
		
		In the following, we present an in-depth derivation of the excitonic Bloch equations, Eqs.~\eqref{eq:chi-3-pol} and \eqref{eq:neg-trion-dyn}.
		To that end, the used carrier Hamiltonian is given in Sec.~\ref{subsec:Hamiltonian}, before deriving Eq.~\eqref{eq:chi-3-pol} in Sec.~\ref{subsec:e-h-pairs} and subsequently deriving Eq.~\eqref{eq:neg-trion-dyn} in Sec.~\ref{subsec:el-herl}.
		
		\subsection{Hamiltonian} \label{subsec:Hamiltonian}

		The carrier Hamiltonian $H$ in the rotating frame approximation \cite{scully1997quantum} is given by \cite{haug2009quantum,mahan2013many}:
		\begin{eqnarray}
			H & = & \sum_{\zeta_1,\textbf{\textit{k}}_1} \varepsilon^{\zeta_1}_{e,\textbf{\textit{k}}_1} \ c^\dag_{\zeta_1,\textbf{\textit{k}}_1} c^{\phantom{\dagger}}_{\zeta_1,\textbf{\textit{k}}_1}
			+ \sum_{\zeta_1,\textbf{\textit{k}}_1} \varepsilon^{\zeta_1}_{h,\textbf{\textit{k}}_1} \ v^\dag_{\zeta_1,\textbf{\textit{k}}_1} v^{\phantom{\dagger}}_{\zeta_1,\textbf{\textit{k}}_1} \notag \\
			& & - \sum_{\zeta_1,\textbf{\textit{k}}_1} \left(d\,^{\zeta_1,\sigma_j}_{c,v\vphantom{\textbf{\textit{k}}_1}} \ \tilde{E}_T^{\sigma_j}(t) \ e^{-i \omega_0 t} \ c_{\zeta_1,\textbf{\textit{k}}_1}^\dag v_{\zeta_1,\textbf{\textit{k}}_1}^{\phantom{\dagger}}
			+ \big[ d\,^{\zeta_1,\sigma_j}_{c,v\vphantom{\textbf{\textit{k}}_1}} \ \tilde{E}_T^{\sigma_j}(t) \ e^{-i \omega_0 t} \big]^* \ v_{\zeta_1,\textbf{\textit{k}}_1}^\dag c_{\zeta_1,\textbf{\textit{k}}_1}^{\phantom{\dagger}}\right) \notag \\
			& & + \frac{1}{2} \sum_{\substack{\zeta_1,\zeta_2 \\ \textbf{\textit{k}}_1,\textbf{\textit{k}}_2,\textbf{\textit{Q}}}} W_{\textbf{\textit{Q}}}  \left(c^\dag_{\zeta_1,\textbf{\textit{k}}_1+\textbf{\textit{Q}}} c^\dag_{\zeta_2,\textbf{\textit{k}}_2-\textbf{\textit{Q}}} c^{\phantom{\dagger}}_{\zeta_2,\textbf{\textit{k}}_2} c^{\phantom{\dagger}}_{\zeta_1,\textbf{\textit{k}}_1} + v^\dag_{\zeta_1,\textbf{\textit{k}}_1+\textbf{\textit{Q}}} v^\dag_{\zeta_2,\textbf{\textit{k}}_2-\textbf{\textit{Q}}} v^{\phantom{\dagger}}_{\zeta_2,\textbf{\textit{k}}_2} v^{\phantom{\dagger}}_{\zeta_1,\textbf{\textit{k}}_1}
			+ 2 \, c^\dag_{\zeta_1,\textbf{\textit{k}}_1+\textbf{\textit{Q}}} v^\dag_{\zeta_2,\textbf{\textit{k}}_2-\textbf{\textit{Q}}} v^{\phantom{\dagger}}_{\zeta_2,\textbf{\textit{k}}_2} c^{\phantom{\dagger}}_{\zeta_1,\textbf{\textit{k}}_1} \right) \label{eq:light-Hamiltonian-electron} . \arxiv
		\end{eqnarray}
		The first line of Eq.~\eqref{eq:light-Hamiltonian-electron} characterizes non-interacting carriers.
		The electron and hole dispersion $\varepsilon^{\zeta}_{e/h,\textbf{\textit{k}}} = \varepsilon\,^{\zeta}_g/2+{\hbar^2\textbf{\textit{k}}^2}/{(2m_{e/h})}$ are described in an effective mass approximation and include the band gap energy $\varepsilon\,^{\zeta}_g$ between conduction and valence band edges as well as the effective electron or hole mass $m_{e/h}$.
		The second line of Eq.~\eqref{eq:light-Hamiltonian-electron} describes interband transitions of carriers between valence and conduction bands due to a $\sigma_j$ circularly polarized light field ($\sigma_j=\sigma_+$, $\sigma_-$) at the monolayer position.
		The interband dipole transition element $d\,^{\zeta_1,\sigma_j}_{c,v}$ is defined in Eq.~\eqref{eq:dipole-ele} and $\tilde{{E}}^{\sigma_j}_{T}(t)$ denotes the envelope of the light field with optical frequency $\omega_0$.
		The last line of Eq.~\eqref{eq:light-Hamiltonian-electron} represents electron-electron, hole-hole, and electron-hole Coulomb interactions with the screened Coulomb potential $W_{\textbf{\textit{Q}}} = V_{\textbf{\textit{Q}}}/\varepsilon_{\textbf{\textit{Q}}}$ provided in Appendix~\ref{app:screen}.

\subsection{Interband Transitions} \label{subsec:e-h-pairs}

		The time evolution of considered observables is determined by the Heisenberg equations of motion: $\partial_t \ \cdot \ = \frac{i}{\hbar} [H, \ \cdot \ ]$.
		Using the carrier Hamiltonian, Eq.~\eqref{eq:light-Hamiltonian-electron}, the dynamics of electron-hole pairs $c^\dag_{\zeta_1,\textbf{\textit{k}}_1} v^{\phantom{\dagger}}_{\zeta_1,\textbf{\textit{k}}_1}$ reads:
		\begin{eqnarray}
			& & \partial_t \ c^\dag_{\zeta_1,\textbf{\textit{k}}_1} v^{\phantom{\dagger}}_{\zeta_1,\textbf{\textit{k}}_1} \notag \\
			& & = \frac{i}{\hbar} \left(\varepsilon\,^{\zeta_1}_g+\frac{\hbar^2\textbf{\textit{k}}_1^2}{2\mu}\right) c^\dag_{\zeta_1,\textbf{\textit{k}}_1} v^{\phantom{\dagger}}_{\zeta_1,\textbf{\textit{k}}_1} 
			- \frac{i}{\hbar} \big[ d\,^{\zeta_1,\sigma_j}_{c,v\vphantom{\textbf{\textit{k}}_1}} \ \tilde{E}_T^{\sigma_j}(t) \ e^{-i \omega_0 t} \big]^* \big(1 - c^\dag_{\zeta_1,\textbf{\textit{k}}_1} c^{\phantom{\dagger}}_{\zeta_1,\textbf{\textit{k}}_1} - v^{\phantom{\dagger}}_{\zeta_1,\textbf{\textit{k}}_1} v^\dag_{\zeta_1,\textbf{\textit{k}}_1} \big) \notag \\
			& & \hspace{4mm} - \frac{i}{\hbar} \sum_{\textbf{\textit{k}}_2} W_{\textbf{\textit{k}}_1-\textbf{\textit{k}}_2} \ c^\dag_{\zeta_1,\textbf{\textit{k}}_2} v^{\phantom{\dagger}}_{\zeta_1,\textbf{\textit{k}}_2}
			+ \frac{i}{\hbar} \sum_{\zeta_2,\textbf{\textit{k}}_2,\textbf{\textit{Q}}} W_{\textbf{\textit{Q}}} \left(c^\dag_{\zeta_1,\textbf{\textit{k}}_1+\textbf{\textit{Q}}} v^{\phantom{\dagger}}_{\zeta_1,\textbf{\textit{k}}_1} c^\dag_{\zeta_2,\textbf{\textit{k}}_2-\textbf{\textit{Q}}} c^{\phantom{\dagger}}_{\zeta_2,\textbf{\textit{k}}_2}
			- c^\dag_{\zeta_1,\textbf{\textit{k}}_1} v^{\phantom{\dagger}}_{\zeta_1,\textbf{\textit{k}}_1+\textbf{\textit{Q}}} c^\dag_{\zeta_2,\textbf{\textit{k}}_2+\textbf{\textit{Q}}} c^{\phantom{\dagger}}_{\zeta_2,\textbf{\textit{k}}_2}\right) \notag \\
			& & \hspace{4mm} + \frac{i}{\hbar} \sum_{\zeta_2,\textbf{\textit{k}}_2,\textbf{\textit{Q}}} W_{\textbf{\textit{Q}}} \left(c^\dag_{\zeta_1,\textbf{\textit{k}}_1} v^{\phantom{\dagger}}_{\zeta_1,\textbf{\textit{k}}_1+\textbf{\textit{Q}}} v^{\phantom{\dagger}}_{\zeta_2,\textbf{\textit{k}}_2} v^\dag_{\zeta_2,\textbf{\textit{k}}_2+\textbf{\textit{Q}}}
			- c^\dag_{\zeta_1,\textbf{\textit{k}}_1+\textbf{\textit{Q}}} v^{\phantom{\dagger}}_{\zeta_1,\textbf{\textit{k}}_1} v^{\phantom{\dagger}}_{\zeta_2,\textbf{\textit{k}}_2} v^\dag_{\zeta_2,\textbf{\textit{k}}_2-\textbf{\textit{Q}}}\right)
			. \label{eq:e-h-pair-un}
		\end{eqnarray}
		The first term on the right-hand side of Eq.~\eqref{eq:e-h-pair-un} describes the free motion depending on the reduced mass $\mu = m_{e}m_{h}/(m_{e}+m_{h})$.
		The second contribution to the right-hand side of Eq.~\eqref{eq:e-h-pair-un} represents the light-matter interaction.
		The last three terms of Eq.~\eqref{eq:e-h-pair-un} characterize Coulomb interactions which introduce a quantum mechanical hierarchy problem due to the coupling of two-operator electron-hole pairs to four-operator terms characterized by the last four contributions to the right-hand side of Eq.~\eqref{eq:e-h-pair-un}.
		The arising hierarchy problem is treated by exploiting a cluster expansion scheme \cite{fricke1996transport}.
		Thus, the dynamics of the interband transitions ${\big\langle} c^\dag_{\zeta_1,\textbf{\textit{k}}_1} v^{\phantom{\dagger}}_{\zeta_1,\textbf{\textit{k}}_1} {\big\rangle}$ becomes:
		\begin{eqnarray}
			& & \left[\partial_t - \frac{i}{\hbar} \left(\varepsilon\,^{\zeta_1}_g+\frac{\hbar^2\textbf{\textit{k}}_1^2}{2\mu} \right)\right]  {\big\langle} c^\dag_{\zeta_1,\textbf{\textit{k}}_1} v^{\phantom{\dagger}}_{\zeta_1,\textbf{\textit{k}}_1} {\big\rangle} \notag \\
			& & =  - \frac{i}{\hbar} \big[ d\,^{\zeta_1,\sigma_j}_{c,v\vphantom{\textbf{\textit{k}}_1}} \ \tilde{E}_T^{\sigma_j}(t) \ e^{-i \omega_0 t} \big]^* \left(1 - f\,^{\zeta_1}_{e,\textbf{\textit{k}}_1} - f\,^{\zeta_1}_{h,\textbf{\textit{k}}_1} \right)
			- \frac{i}{\hbar} \sum_{\textbf{\textit{k}}_2} W_{\textbf{\textit{k}}_1-\textbf{\textit{k}}_2} \ {\big\langle} c^\dag_{\zeta_1,\textbf{\textit{k}}_2} v^{\phantom{\dagger}}_{\zeta_1,\textbf{\textit{k}}_2} {\big\rangle} \notag \\
			& & \hspace{4mm} - \frac{i}{\hbar} \sum_{\textbf{\textit{k}}_2} W_{\textbf{\textit{k}}_1-\textbf{\textit{k}}_2} \ {\big\langle} c^\dag_{\zeta_1,\textbf{\textit{k}}_1} v^{\phantom{\dagger}}_{\zeta_1,\textbf{\textit{k}}_1} {\big\rangle} \left(f\,^{\zeta_1}_{e,\textbf{\textit{k}}_2} + f\,^{\zeta_1}_{h,\textbf{\textit{k}}_2}\right)
			+ \frac{i}{\hbar} \sum_{\textbf{\textit{k}}_2} W_{\textbf{\textit{k}}_1-\textbf{\textit{k}}_2} \ {\big\langle} c^\dag_{\zeta_1,\textbf{\textit{k}}_2} v^{\phantom{\dagger}}_{\zeta_1,\textbf{\textit{k}}_2} {\big\rangle} \left(f\,^{\zeta_1}_{e,\textbf{\textit{k}}_1} + f\,^{\zeta_1}_{h,\textbf{\textit{k}}_1}\right) \notag \\		
			& & \hspace{4mm} + \frac{i}{\hbar} \sum_{\zeta_2,\textbf{\textit{k}}_2,\textbf{\textit{Q}}} W_{\textbf{\textit{Q}}} \left({\big\langle} c^\dag_{\zeta_1,\textbf{\textit{k}}_1+\textbf{\textit{Q}}} v^{\phantom{\dagger}}_{\zeta_1,\textbf{\textit{k}}_1} c^\dag_{\zeta_2,\textbf{\textit{k}}_2-\textbf{\textit{Q}}} c^{\phantom{\dagger}}_{\zeta_2,\textbf{\textit{k}}_2} {\big\rangle}^{c}
			- {\big\langle} c^\dag_{\zeta_1,\textbf{\textit{k}}_1} v^{\phantom{\dagger}}_{\zeta_1,\textbf{\textit{k}}_1+\textbf{\textit{Q}}} c^\dag_{\zeta_2,\textbf{\textit{k}}_2+\textbf{\textit{Q}}} c^{\phantom{\dagger}}_{\zeta_2,\textbf{\textit{k}}_2} {\big\rangle}^{c}\right) \notag \\
			& & \hspace{4mm} + \frac{i}{\hbar} \sum_{\zeta_2,\textbf{\textit{k}}_2,\textbf{\textit{Q}}} W_{\textbf{\textit{Q}}} \left({\big\langle} c^\dag_{\zeta_1,\textbf{\textit{k}}_1} v^{\phantom{\dagger}}_{\zeta_1,\textbf{\textit{k}}_1+\textbf{\textit{Q}}} v^{\phantom{\dagger}}_{\zeta_2,\textbf{\textit{k}}_2} v^\dag_{\zeta_2,\textbf{\textit{k}}_2+\textbf{\textit{Q}}} {\big\rangle}^{c}
			- c^\dag_{\zeta_1,\textbf{\textit{k}}_1+\textbf{\textit{Q}}} v^{\phantom{\dagger}}_{\zeta_1,\textbf{\textit{k}}_1} v^{\phantom{\dagger}}_{\zeta_2,\textbf{\textit{k}}_2} v^\dag_{\zeta_2,\textbf{\textit{k}}_2-\textbf{\textit{Q}}} {\big\rangle}^{c}\right) . \label{eq:e-h-pair-fa}
		\end{eqnarray}
		Here, electron occupations are characterized by residual electron densities $f\,^{\zeta_1}_{e,\textbf{\textit{k}}_1} = {\big\langle} c^\dag_{\zeta_1,\textbf{\textit{k}}_1} c^{\phantom{\dagger}}_{\zeta_1,\textbf{\textit{k}}_1} {\big\rangle}$ and hole occupations are determined by residual hole densities $f\,^{\zeta_1}_{h,\textbf{\textit{k}}_1} = {\big\langle} v^{\phantom{\dagger}}_{\zeta_1,\textbf{\textit{k}}_1} v^\dag_{\zeta_1,\textbf{\textit{k}}_1} {\big\rangle}$.
		The electron-density-assisted transitions ${\big\langle} c^\dag_{\zeta_1,\textbf{\textit{k}}_1+\textbf{\textit{Q}}} v^{\phantom{\dagger}}_{\zeta_1,\textbf{\textit{k}}_1} c^\dag_{\zeta_2,\textbf{\textit{k}}_2-\textbf{\textit{Q}}} c^{\phantom{\dagger}}_{\zeta_2,\textbf{\textit{k}}_2} {\big\rangle}^{c}$ are defined in Eq.~\eqref{eq:fat-e} and the hole-density-assisted transitions ${\big\langle} c^\dag_{\zeta_1,\textbf{\textit{k}}_1+\textbf{\textit{Q}}} v^{\phantom{\dagger}}_{\zeta_1,\textbf{\textit{k}}_1} v^{\phantom{\dagger}}_{\zeta_2,\textbf{\textit{k}}_2+\textbf{\textit{Q}}} v^\dag_{\zeta_2,\textbf{\textit{k}}_2} {\big\rangle}^{c}$ are defined in Eq.~\eqref{eq:fat-h}.
		The second term on the right-hand side of Eq.~\eqref{eq:e-h-pair-fa} introduces a Coulomb-mediated coupling among interband transitions ${\big\langle} c^\dag_{\zeta_1,\textbf{\textit{k}}_1} v^{\phantom{\dagger}}_{\zeta_1,\textbf{\textit{k}}_1} {\big\rangle}$ and ${\big\langle} c^\dag_{\zeta_1,\textbf{\textit{k}}_2} v^{\phantom{\dagger}}_{\zeta_1,\textbf{\textit{k}}_2} {\big\rangle}$ with different wave vectors $\textbf{\textit{k}}_1$ and $\textbf{\textit{k}}_2$.
		After identifying the Wannier equation, Eq.~\eqref{eq:Wannier-Gl}, the interband transitions ${\big\langle} c^\dag_{\zeta_1,\textbf{\textit{k}}_1} v^{\phantom{\dagger}}_{\zeta_1,\textbf{\textit{k}}_1} {\big\rangle}$ can be expanded in terms of a complete set of exciton wave functions $\varphi^R\,^{\zeta_1}_{\nu_2,\textbf{\textit{k}}_1}$ and exciton transitions $P\,^{\zeta_1}_{\nu_2}$ according to Eq.~\eqref{eq:definition-exciton}.
		The third and fourth term on the right-hand side of Eq.~\eqref{eq:e-h-pair-fa} reduce the band gap energy and exciton binding energy, respectively.
		Equation~\eqref{eq:e-h-pair-fa} is now multiplied by $\frac{1}{\mathcal{A}} \sum_{\textbf{\textit{k}}_1} \varphi^L\,^{\zeta_1}_{\nu_1,\textbf{\textit{k}}_1}$ and the normalization of exciton wave functions, Eq.~\eqref{eq:normierung-x}, can be employed.
		Using the definitions of the Rabi frequency $\Omega\,^{\zeta_1,\sigma_j}_{\nu_1,\textbf{\textit{k}}_1}$, Eq.~\eqref{eq:Rabi-freq}, and the Hartree--Fock Coulomb matrix element $\hat{W}\,^{\zeta_1}_{H\text{-}F,\nu_1,\nu_2,\textbf{\textit{Q}}} $, Eq.~\eqref{eq:mat-el-hf}, the dynamics of the exciton transitions $P\,^{\zeta_1}_{\nu_1}$ becomes:
		\begin{eqnarray}
			& & \left(\partial_t - \frac{i}{\hbar} \epsilon\,^{\zeta_1}_{x,\nu_1}\right) P\,^{\zeta_1}_{\nu_1} \notag \\
			& & = - \frac{i}{\mathcal{A}} \sum_{\textbf{\textit{k}}_1} \Omega\,^{\zeta_1,\sigma_j}_{\nu_1,\textbf{\textit{k}}_1} \left(1 - f\,^{\zeta_1}_{e,\textbf{\textit{k}}_1} - f\,^{\zeta_1}_{h,\textbf{\textit{k}}_1} \right)
			+ \frac{i}{\hbar\mathcal{A}} \sum_{\nu_2,\textbf{\textit{k}}_1} \hat{W}\,^{\zeta_1}_{H\text{-}F,\nu_2,\nu_1,\textbf{\textit{k}}_1} \left(f\,^{\zeta_1}_{e,\textbf{\textit{k}}_1} + f\,^{\zeta_1}_{h,\textbf{\textit{k}}_1}\right)  P\,^{\zeta_1}_{\nu_2} \notag \\
			& & \hspace{3.9mm} + \frac{i}{\hbar\mathcal{A}} \sum_{\zeta_2,\textbf{\textit{k}}_1,\textbf{\textit{k}}_2,\textbf{\textit{Q}}} W_{\textbf{\textit{Q}}} \left(\varphi^L\,^{\zeta_1}_{\nu_1,\textbf{\textit{k}}_1} - \varphi^L\,^{\zeta_1}_{\nu_1,\textbf{\textit{k}}_1+\textbf{\textit{Q}}}\right)
			\Big({\big\langle} c^\dag_{\zeta_1,\textbf{\textit{k}}_1+\textbf{\textit{Q}}} v^{\phantom{\dagger}}_{\zeta_1,\textbf{\textit{k}}_1} c^\dag_{\zeta_2,\textbf{\textit{k}}_2-\textbf{\textit{Q}}} c^{\phantom{\dagger}}_{\zeta_2,\textbf{\textit{k}}_2} {\big\rangle}^{c} 
			- {\big\langle} c^\dag_{\zeta_1,\textbf{\textit{k}}_1+\textbf{\textit{Q}}} v^{\phantom{\dagger}}_{\zeta_1,\textbf{\textit{k}}_1} v^{\phantom{\dagger}}_{\zeta_2,\textbf{\textit{k}}_2+\textbf{\textit{Q}}} v^\dag_{\zeta_2,\textbf{\textit{k}}_2} {\big\rangle}^{c}\Big) . \arxiv \label{eq:e-h-pair-fa2}
		\end{eqnarray}
		Finally, the electron- and hole-density-assisted transitions on the right-hand side of Eq.~\eqref{eq:e-h-pair-fa2} can be substituted according to Eqs.~\eqref{eq:electron-screened} and \eqref{eq:hole-screened} which finally leads to Eq.~\eqref{eq:chi-3-pol}.

\subsection{Electron-Density-Assisted Transitions} \label{subsec:el-herl}

			The Heisenberg equation of motion of electron-density-assisted transitions $c^\dag_{\zeta_1,\textbf{\textit{k}}_1+\textbf{\textit{Q}}} v^{\phantom{\dagger}}_{\zeta_2,\textbf{\textit{k}}_1} c^\dag_{\zeta_3,\textbf{\textit{k}}_2-\textbf{\textit{Q}}} c^{\phantom{\dagger}}_{\zeta_4,\textbf{\textit{k}}_2}$ is calculated using the carrier Hamiltonian, Eq.~\eqref{eq:light-Hamiltonian-electron}:
			\begin{eqnarray}
			& & \partial_t \ c^\dag_{\zeta_1,\textbf{\textit{k}}_1+\textbf{\textit{Q}}} v^{\phantom{\dagger}}_{\zeta_2,\textbf{\textit{k}}_1} c^\dag_{\zeta_3,\textbf{\textit{k}}_2-\textbf{\textit{Q}}} c^{\phantom{\dagger}}_{\zeta_4,\textbf{\textit{k}}_2} \notag \\
			& & = \frac{i}{\hbar} \left[\varepsilon\,^{\zeta_1}_g + \frac{\hbar^2}{2}\left(\frac{\textbf{\textit{k}}_1^2}{\mu} + \frac{2\textbf{\textit{Q}}^2+2\textbf{\textit{k}}_1\cdot\textbf{\textit{Q}}-2\textbf{\textit{k}}_2\cdot\textbf{\textit{Q}}}{m^{e}}\right) \right] c^\dag_{\zeta_1,\textbf{\textit{k}}_1+\textbf{\textit{Q}}} v^{\phantom{\dagger}}_{\zeta_2,\textbf{\textit{k}}_1} c^\dag_{\zeta_3,\textbf{\textit{k}}_2-\textbf{\textit{Q}}} c^{\phantom{\dagger}}_{\zeta_4,\textbf{\textit{k}}_2} \notag \\
			& & \hspace{3.9mm} - \frac{i}{\hbar} \delta_{\zeta_1,\zeta_2} \delta_{\textbf{\textit{Q}},\textbf{0}} \big[ d\,^{\zeta_1,\sigma_j}_{c,v\vphantom{\textbf{\textit{k}}_1}} \ \tilde{E}_T^{\sigma_j}(t) \ e^{-i \omega_0 t} \big]^* \ c^\dag_{\zeta_3,\textbf{\textit{k}}_2} c^{\phantom{\dagger}}_{\zeta_4,\textbf{\textit{k}}_2}
			+ \frac{i}{\hbar} \delta_{\zeta_2,\zeta_3}\delta_{\textbf{\textit{k}}_1+\textbf{\textit{Q}},\textbf{\textit{k}}_2} \big[ d\,^{\zeta_2,\sigma_j}_{c,v\vphantom{\textbf{\textit{k}}_1}} \ \tilde{E}_T^{\sigma_j}(t) \ e^{-i \omega_0 t} \big]^* \ c^\dag_{\zeta_1,\textbf{\textit{k}}_1+\textbf{\textit{Q}}} c^{\phantom{\dagger}}_{\zeta_4,\textbf{\textit{k}}_1+\textbf{\textit{Q}}} \notag \\
			& & \hspace{3.9mm} - \frac{i}{\hbar} \big[ d\,^{\zeta_2,\sigma_j}_{c,v\vphantom{\textbf{\textit{k}}_1}} \ \tilde{E}_T^{\sigma_j}(t) \ e^{-i \omega_0 t} \big]^* \ c^\dag_{\zeta_1,\textbf{\textit{k}}_1+\textbf{\textit{Q}}} c^\dag_{\zeta_3,\textbf{\textit{k}}_2-\textbf{\textit{Q}}} c^{\phantom{\dagger}}_{\zeta_2,\textbf{\textit{k}}_1} c^{\phantom{\dagger}}_{\zeta_4,\textbf{\textit{k}}_2}
			+ \frac{i}{\hbar} \big[ d\,^{\zeta_1,\sigma_j}_{c,v\vphantom{\textbf{\textit{k}}_1}} \ \tilde{E}_T^{\sigma_j}(t) \ e^{-i \omega_0 t} \big]^* \ c^\dag_{\zeta_3,\textbf{\textit{k}}_2-\textbf{\textit{Q}}} v^{\phantom{\dagger}}_{\zeta_2,\textbf{\textit{k}}_1} v^\dag_{\zeta_1,\textbf{\textit{k}}_1+\textbf{\textit{Q}}} c^{\phantom{\dagger}}_{\zeta_4,\textbf{\textit{k}}_2} \notag \\
			& & \hspace{3.9mm} - \frac{i}{\hbar} \big[ d\,^{\zeta_3,\sigma_j}_{c,v\vphantom{\textbf{\textit{k}}_1}} \ \tilde{E}_T^{\sigma_j}(t) \ e^{-i \omega_0 t} \big]^* \ c^\dag_{\zeta_1,\textbf{\textit{k}}_1+\textbf{\textit{Q}}} v^{\phantom{\dagger}}_{\zeta_2,\textbf{\textit{k}}_1} v^\dag_{\zeta_3,\textbf{\textit{k}}_2-\textbf{\textit{Q}}} c^{\phantom{\dagger}}_{\zeta_4,\textbf{\textit{k}}_2}
			+ \frac{i}{\hbar} \ d\,^{\zeta_4,\sigma_j}_{c,v\vphantom{\textbf{\textit{k}}_1}} \ \tilde{E}_T^{\sigma_j}(t) \ e^{-i \omega_0 t} \ c^\dag_{\zeta_1,\textbf{\textit{k}}_1+\textbf{\textit{Q}}} v^{\phantom{\dagger}}_{\zeta_2,\textbf{\textit{k}}_1} c^\dag_{\zeta_3,\textbf{\textit{k}}_2-\textbf{\textit{Q}}} v^{\phantom{\dagger}}_{\zeta_4,\textbf{\textit{k}}_2} \notag \\
			& & \hspace{3.9mm} + \frac{i}{\hbar} \sum_{\textbf{\textit{Q}}'} W_{\textbf{\textit{Q}}'} \ \Big(c^\dag_{\zeta_1,\textbf{\textit{k}}_1+\textbf{\textit{Q}}+\textbf{\textit{Q}}'} v^{\phantom{\dagger}}_{\zeta_2,\textbf{\textit{k}}_1} c^\dag_{\zeta_3,\textbf{\textit{k}}_2-\textbf{\textit{Q}}-\textbf{\textit{Q}}'} c^{\phantom{\dagger}}_{\zeta_4,\textbf{\textit{k}}_2}
			- c^\dag_{\zeta_1,\textbf{\textit{k}}_1+\textbf{\textit{Q}}+\textbf{\textit{Q}}'} v^{\phantom{\dagger}}_{\zeta_2,\textbf{\textit{k}}_1+\textbf{\textit{Q}}'} c^\dag_{\zeta_3,\textbf{\textit{k}}_2-\textbf{\textit{Q}}} c^{\phantom{\dagger}}_{\zeta_4,\textbf{\textit{k}}_2} \notag \\
			& & \hspace{25.4mm} - c^\dag_{\zeta_1,\textbf{\textit{k}}_1+\textbf{\textit{Q}}} v^{\phantom{\dagger}}_{\zeta_2,\textbf{\textit{k}}_1+\textbf{\textit{Q}}'} c^\dag_{\zeta_3,\textbf{\textit{k}}_2-\textbf{\textit{Q}}+\textbf{\textit{Q}}'} c^{\phantom{\dagger}}_{\zeta_4,\textbf{\textit{k}}_2} 
			+ c^\dag_{\zeta_1,\textbf{\textit{k}}_1+\textbf{\textit{Q}}+\textbf{\textit{Q}}'} v^{\phantom{\dagger}}_{\zeta_2,\textbf{\textit{k}}_1} c^\dag_{\zeta_3,\textbf{\textit{k}}_2-\textbf{\textit{Q}}} c^\dag_{\zeta_5,\textbf{\textit{k}}_3-\textbf{\textit{Q}}'} c^{\phantom{\dagger}}_{\zeta_5,\textbf{\textit{k}}_3} c^{\phantom{\dagger}}_{\zeta_4,\textbf{\textit{k}}_2} \notag \\
			& & \hspace{25.4mm} + c^\dag_{\zeta_1,\textbf{\textit{k}}_1+\textbf{\textit{Q}}} v^{\phantom{\dagger}}_{\zeta_2,\textbf{\textit{k}}_1} c^\dag_{\zeta_3,\textbf{\textit{k}}_2-\textbf{\textit{Q}}+\textbf{\textit{Q}}'} c^\dag_{\zeta_5,\textbf{\textit{k}}_3-\textbf{\textit{Q}}'} c^{\phantom{\dagger}}_{\zeta_5,\textbf{\textit{k}}_3} c^{\phantom{\dagger}}_{\zeta_4,\textbf{\textit{k}}_2} 
			- c^\dag_{\zeta_1,\textbf{\textit{k}}_1+\textbf{\textit{Q}}} v^{\phantom{\dagger}}_{\zeta_2,\textbf{\textit{k}}_1} c^\dag_{\zeta_3,\textbf{\textit{k}}_2-\textbf{\textit{Q}}} c^\dag_{\zeta_5,\textbf{\textit{k}}_3+\textbf{\textit{Q}}'} c^{\phantom{\dagger}}_{\zeta_5,\textbf{\textit{k}}_3} c^{\phantom{\dagger}}_{\zeta_4,\textbf{\textit{k}}_2+\textbf{\textit{Q}}'} \notag \\
			& & \hspace{25.4mm} - c^\dag_{\zeta_1,\textbf{\textit{k}}_1+\textbf{\textit{Q}}} v^{\phantom{\dagger}}_{\zeta_2,\textbf{\textit{k}}_1+\textbf{\textit{Q}}'} c^\dag_{\zeta_3,\textbf{\textit{k}}_2-\textbf{\textit{Q}}} c^\dag_{\zeta_5,\textbf{\textit{k}}_3+\textbf{\textit{Q}}'} c^{\phantom{\dagger}}_{\zeta_5,\textbf{\textit{k}}_3} c^{\phantom{\dagger}}_{\zeta_4,\textbf{\textit{k}}_2}
			+ c^\dag_{\zeta_1,\textbf{\textit{k}}_1+\textbf{\textit{Q}}} v^{\phantom{\dagger}}_{\zeta_2,\textbf{\textit{k}}_1+\textbf{\textit{Q}}'} c^\dag_{\zeta_3,\textbf{\textit{k}}_2-\textbf{\textit{Q}}} v^{\phantom{\dagger}}_{\zeta_5,\textbf{\textit{k}}_3} v^\dag_{\zeta_5,\textbf{\textit{k}}_3+\textbf{\textit{Q}}'} c^{\phantom{\dagger}}_{\zeta_4,\textbf{\textit{k}}_2} \notag \\
			& & \hspace{25.4mm} - c^\dag_{\zeta_1,\textbf{\textit{k}}_1+\textbf{\textit{Q}}+\textbf{\textit{Q}}'} v^{\phantom{\dagger}}_{\zeta_2,\textbf{\textit{k}}_1} c^\dag_{\zeta_3,\textbf{\textit{k}}_2-\textbf{\textit{Q}}} v^{\phantom{\dagger}}_{\zeta_5,\textbf{\textit{k}}_3} v^\dag_{\zeta_5,\textbf{\textit{k}}_3-\textbf{\textit{Q}}'} c^{\phantom{\dagger}}_{\zeta_4,\textbf{\textit{k}}_2}
			- c^\dag_{\zeta_1,\textbf{\textit{k}}_1+\textbf{\textit{Q}}} v^{\phantom{\dagger}}_{\zeta_2,\textbf{\textit{k}}_1} c^\dag_{\zeta_3,\textbf{\textit{k}}_2-\textbf{\textit{Q}}+\textbf{\textit{Q}}'} v^{\phantom{\dagger}}_{\zeta_5,\textbf{\textit{k}}_3} v^\dag_{\zeta_5,\textbf{\textit{k}}_3-\textbf{\textit{Q}}'} c^{\phantom{\dagger}}_{\zeta_4,\textbf{\textit{k}}_2} \notag \\
			& & \hspace{25.4mm} + c^\dag_{\zeta_1,\textbf{\textit{k}}_1+\textbf{\textit{Q}}} v^{\phantom{\dagger}}_{\zeta_2,\textbf{\textit{k}}_1} c^\dag_{\zeta_3,\textbf{\textit{k}}_2-\textbf{\textit{Q}}} v^{\phantom{\dagger}}_{\zeta_5,\textbf{\textit{k}}_3} v^\dag_{\zeta_5,\textbf{\textit{k}}_3-\textbf{\textit{Q}}'} c^{\phantom{\dagger}}_{\zeta_4,\textbf{\textit{k}}_2-\textbf{\textit{Q}}'} \Big) \label{eq:ass-tran-1} .
			\end{eqnarray}
			The first term on the right-hand side of Eq.~\eqref{eq:ass-tran-1} describes the free motion.
			The second to seventh contributions to the right-hand side of Eq.~\eqref{eq:ass-tran-1} characterize the light-matter coupling.
			Here, the term associated with the four-operator term $c^\dag_{\zeta_1,\textbf{\textit{k}}_1+\textbf{\textit{Q}}} c^\dag_{\zeta_3,\textbf{\textit{k}}_2-\textbf{\textit{Q}}} c^{\phantom{\dagger}}_{\zeta_2,\textbf{\textit{k}}_1} c^{\phantom{\dagger}}_{\zeta_4,\textbf{\textit{k}}_2}$ can be either represented by quadratic doping densities or an exciton density \cite{katsch2018theory} and will be subsequently neglected by truncating the dynamics to linear doping densities and the linear optical response solely.
			The fifth to seventh contributions to the right-hand side of Eq.~\eqref{eq:ass-tran-1} represent nonlinear $\chi^{(3)}$ terms \cite{axt1994dynamics,axt1994role,lindberg1994chi} which go beyond the linear optical response, also referred to as $\chi^{(1)}$ regime, and are neglected.
			Coulomb interactions, described by the last term on the right-hand side of Eq.~\eqref{eq:ass-tran-1}, lead to a quantum mechanical hierarchy problem due to the coupling of the electron-density-assisted transitions, represented by four-operator terms, to six-operator terms.
			The arising hierarchy problem is also truncated to the linear optical response and linear doping densities, where the six-operator terms are neglected and only the four-operator terms are considered \cite{esser2001theory}.
			Next, a cluster expansion is exploited to express Eq.~\eqref{eq:ass-tran-1} in terms of interband transitions ${\big\langle} c^\dag_{\zeta_1,\textbf{\textit{k}}_1} v^{\phantom{\dagger}}_{\zeta_1,\textbf{\textit{k}}_1} {\big\rangle}$, electron densities $f\,^{\zeta_1}_{e,\textbf{\textit{k}}_1} = {\big\langle} c^\dag_{\zeta_1,\textbf{\textit{k}}_1} c^{\phantom{\dagger}}_{\zeta_1,\textbf{\textit{k}}_1} {\big\rangle}$, and electron-density-assisted transitions ${\big\langle} c^\dag_{\zeta_1,\textbf{\textit{k}}_1+\textbf{\textit{Q}}} v^{\phantom{\dagger}}_{\zeta_1,\textbf{\textit{k}}_1} c^\dag_{\zeta_2,\textbf{\textit{k}}_2-\textbf{\textit{Q}}} c^{\phantom{\dagger}}_{\zeta_2,\textbf{\textit{k}}_2} {\big\rangle}^{c}$, as introduced in Eq.~\eqref{eq:fat-e}:
			\begin{eqnarray}
			& & \left\lbrace \partial_t - \frac{i}{\hbar} \left[\varepsilon\,^{\zeta_1}_g + \frac{\hbar^2}{2}\left(\frac{\textbf{\textit{k}}_1^2}{\mu} + \frac{2\textbf{\textit{Q}}^2+2\textbf{\textit{k}}_1\cdot\textbf{\textit{Q}}-2\textbf{\textit{k}}_2\cdot\textbf{\textit{Q}}}{m^{e}}\right) \right] \right\rbrace {\big\langle} c^\dag_{\zeta_1,\textbf{\textit{k}}_1+\textbf{\textit{Q}}} v^{\phantom{\dagger}}_{\zeta_2,\textbf{\textit{k}}_1} c^\dag_{\zeta_3,\textbf{\textit{k}}_2-\textbf{\textit{Q}}} c^{\phantom{\dagger}}_{\zeta_4,\textbf{\textit{k}}_2} {\big\rangle}^{c} \notag \\
			& & = \frac{i \, \delta_{\zeta_1,\zeta_2} \, \delta_{\zeta_3,\zeta_4}}{\hbar} \ W_{\textbf{\textit{Q}}} \left({\big\langle} c^\dag_{\zeta_1,\textbf{\textit{k}}_1} v^{\phantom{\dagger}}_{\zeta_1,\textbf{\textit{k}}_1} {\big\rangle} - {\big\langle} c^\dag_{\zeta_1,\textbf{\textit{k}}_1+\textbf{\textit{Q}}} v^{\phantom{\dagger}}_{\zeta_1,\textbf{\textit{k}}_1+\textbf{\textit{Q}}} {\big\rangle}\right)f\,^{\zeta_3}_{e,\textbf{\textit{k}}_2} \notag \\
			& & \hspace{3.9mm} - \frac{i \, \delta_{\zeta_1,\zeta_4} \, \delta_{\zeta_2,\zeta_3}}{\hbar} \ W_{\textbf{\textit{k}}_1-\textbf{\textit{k}}_2+\textbf{\textit{Q}}} \left({\big\langle} c^\dag_{\zeta_2,\textbf{\textit{k}}_1} v^{\phantom{\dagger}}_{\zeta_2,\textbf{\textit{k}}_1} {\big\rangle} - {\big\langle} c^\dag_{\zeta_2,\textbf{\textit{k}}_2-\textbf{\textit{Q}}} v^{\phantom{\dagger}}_{\zeta_2,\textbf{\textit{k}}_2-\textbf{\textit{Q}}} {\big\rangle}\right)f\,^{\zeta_1}_{e,\textbf{\textit{k}}_2} \notag \\
			& & \hspace{3.9mm} + \frac{i}{\hbar} \sum_{\textbf{\textit{Q}}'} W_{\textbf{\textit{Q}}'} \ {\big\langle} c^\dag_{\zeta_1,\textbf{\textit{k}}_1+\textbf{\textit{Q}}+\textbf{\textit{Q}}'} v^{\phantom{\dagger}}_{\zeta_2,\textbf{\textit{k}}_1} c^\dag_{\zeta_3,\textbf{\textit{k}}_2-\textbf{\textit{Q}}-\textbf{\textit{Q}}'} c^{\phantom{\dagger}}_{\zeta_4,\textbf{\textit{k}}_2} {\big\rangle}^{c} \notag \\
			& & \hspace{3.9mm} - \frac{i}{\hbar} \sum_{\textbf{\textit{Q}}'} W_{\textbf{\textit{Q}}'} \ {\big\langle} c^\dag_{\zeta_1,\textbf{\textit{k}}_1+\textbf{\textit{Q}}+\textbf{\textit{Q}}'} v^{\phantom{\dagger}}_{\zeta_2,\textbf{\textit{k}}_1+\textbf{\textit{Q}}'} c^\dag_{\zeta_3,\textbf{\textit{k}}_2-\textbf{\textit{Q}}} c^{\phantom{\dagger}}_{\zeta_4,\textbf{\textit{k}}_2} {\big\rangle}^{c} \notag \\
			& & \hspace{3.9mm} - \frac{i}{\hbar} \sum_{\textbf{\textit{Q}}'} W_{\textbf{\textit{Q}}'} \ {\big\langle} c^\dag_{\zeta_1,\textbf{\textit{k}}_1+\textbf{\textit{Q}}} v^{\phantom{\dagger}}_{\zeta_2,\textbf{\textit{k}}_1+\textbf{\textit{Q}}'} c^\dag_{\zeta_3,\textbf{\textit{k}}_2-\textbf{\textit{Q}}+\textbf{\textit{Q}}'} c^{\phantom{\dagger}}_{\zeta_4,\textbf{\textit{k}}_2} {\big\rangle}^{c} .  \label{eq:ass-tran-2}
			\end{eqnarray}
			Even though the free energy on the left-hand side of Eq.~\eqref{eq:ass-tran-2} contains no contributions proportional to $\textbf{\textit{k}}_1\cdot\textbf{\textit{k}}_2$, there appear terms of the form $\textbf{\textit{k}}_1\cdot\textbf{\textit{Q}}$ and $\textbf{\textit{k}}_2\cdot\textbf{\textit{Q}}$.
			Therefore, it is convenient to use the three wave vectors $\textbf{\textit{K}}_1$, $\textbf{\textit{k}}_2$, and $\textbf{\textit{Q}}_2$ defined by $\textbf{\textit{k}}_1 = \textbf{\textit{K}}_1-\beta_{x\text{-}e}\textbf{\textit{k}}_2-\beta_{x}\textbf{\textit{Q}}_2$ and $\textbf{\textit{Q}} = (1-\alpha_{x\text{-}e})\textbf{\textit{k}}_2+\textbf{\textit{Q}}_2$.
			The advantage of the new set of wave vectors $\textbf{\textit{K}}_1$, $\textbf{\textit{k}}_2$, and $\textbf{\textit{Q}}_2$ is that no terms of the form $\textbf{\textit{K}}_1\cdot\textbf{\textit{k}}_2$, $\textbf{\textit{K}}_1\cdot\textbf{\textit{Q}}_2$, or $\textbf{\textit{k}}_2\cdot\textbf{\textit{Q}}_2$ contribute to the free energy anymore.
			As a result, the following relation can be proven by using the definition of the reduced mass $\mu = m_{e}m_{h}/(m_{e}+m_{h})$ as well as the definitions of the ratios $\alpha_{x}$, $\beta_{x}$, $\alpha_{x\text{-}e}$, and $\beta_{x\text{-}e}$ given in Eqs.~\eqref{eq:alpha-beta} and \eqref{eq:alpha-beta-e}:
			\begin{equation}
			\frac{\textbf{\textit{k}}_1^2}{\mu} + \frac{2\textbf{\textit{Q}}^2+2\textbf{\textit{k}}_1\cdot\textbf{\textit{Q}}-2\textbf{\textit{k}}_2\cdot\textbf{\textit{Q}}}{m^{e}} = \frac{\textbf{\textit{K}}_1^2}{\mu} - \textbf{\textit{k}}_2^2\frac{m_e+m_h}{(2m_e+m_h)m_e} +\textbf{\textit{Q}}_2^2\left(\frac{1}{m_e+m_h}+\frac{1}{m_e}\right) .
			\end{equation}
			Thus, writing Eq.~\eqref{eq:ass-tran-2} in terms of the new set of wave vectors $\textbf{\textit{K}}_1$, $\textbf{\textit{k}}_2$, and $\textbf{\textit{Q}}_2$ instead of $\textbf{\textit{k}}_1$, $\textbf{\textit{k}}_2$, and $\textbf{\textit{Q}}$ leads to:
			\begin{eqnarray}
			& & \left\lbrace\partial_t - \frac{i}{\hbar} \left[ \varepsilon\,^{\zeta_1}_g + \frac{\hbar^2\textbf{\textit{K}}_1^2}{2\mu} -\frac{\hbar^2\textbf{\textit{k}}_2^2}{2}\frac{m_e+m_h}{(2m_e+m_h)m_e} +\frac{\hbar^2\textbf{\textit{Q}}_2^2}{2}\left(\frac{1}{m_e+m_h}+\frac{1}{m_e}\right) \right] \right\rbrace \notag \\
			& & \times {\big\langle} c^\dag_{\zeta_1,\textbf{\textit{K}}_1+\alpha_{x\text{-}e}\textbf{\textit{k}}_2+\alpha_{x}\textbf{\textit{Q}}_2} v^{\phantom{\dagger}}_{\zeta_2,\textbf{\textit{K}}_1-\beta_{x\text{-}e}\textbf{\textit{k}}_2-\beta_{x}\textbf{\textit{Q}}_2} c^\dag_{\zeta_3,\alpha_{x\text{-}e}\textbf{\textit{k}}_2-\textbf{\textit{Q}}_2} c^{\phantom{\dagger}}_{\zeta_4,\textbf{\textit{k}}_2} {\big\rangle}^{c} \notag \\
			& & = \frac{i \, \delta_{\zeta_1,\zeta_2} \, \delta_{\zeta_3,\zeta_4}}{\hbar} \ W_{(1-\alpha_{x\text{-}e})\textbf{\textit{k}}_2+\textbf{\textit{Q}}_2} \Big({\big\langle} c^\dag_{\zeta_1,\textbf{\textit{K}}_1-\beta_{x\text{-}e}\textbf{\textit{k}}_2-\beta_{x}\textbf{\textit{Q}}_2} v^{\phantom{\dagger}}_{\zeta_1,\textbf{\textit{K}}_1-\beta_{x\text{-}e}\textbf{\textit{k}}_2-\beta_{x}\textbf{\textit{Q}}_2} {\big\rangle} \notag \\ & & \hspace{49.4mm} - {\big\langle} c^\dag_{\zeta_1,\textbf{\textit{K}}_1+\alpha_{x\text{-}e}\textbf{\textit{k}}_2+\alpha_{x}\textbf{\textit{Q}}_2} v^{\phantom{\dagger}}_{\zeta_1,\textbf{\textit{K}}_1+\alpha_{x\text{-}e}\textbf{\textit{k}}_2+\alpha_{x}\textbf{\textit{Q}}_2} {\big\rangle}\Big)f\,^{\zeta_3}_{e,\textbf{\textit{k}}_2} \notag \\
			& & \hspace{3.9mm} - \frac{i \, \delta_{\zeta_1,\zeta_4} \, \delta_{\zeta_2,\zeta_3}}{\hbar} \ W_{\textbf{\textit{K}}_1-(1-\alpha_{x\text{-}e})\textbf{\textit{k}}_2+\alpha_{x}\textbf{\textit{Q}}_2} \Big({\big\langle} c^\dag_{\zeta_2,\textbf{\textit{K}}_1-\beta_{x\text{-}e}\textbf{\textit{k}}_2-\beta_{x}\textbf{\textit{Q}}_2} v^{\phantom{\dagger}}_{\zeta_2,\textbf{\textit{K}}_1-\beta_{x\text{-}e}\textbf{\textit{k}}_2-\beta_{x}\textbf{\textit{Q}}_2} {\big\rangle} \notag \\ & & \hspace{62.0mm} - {\big\langle} c^\dag_{\zeta_2,\alpha_{x\text{-}e}\textbf{\textit{k}}_2-\textbf{\textit{Q}}_2} v^{\phantom{\dagger}}_{\zeta_2,\alpha_{x\text{-}e}\textbf{\textit{k}}_2-\textbf{\textit{Q}}_2} {\big\rangle}\Big)f\,^{\zeta_1}_{e,\textbf{\textit{k}}_2} \notag \\
			& & \hspace{3.9mm} + \frac{i}{2\hbar} \sum_{\textbf{\textit{Q}}'} W_{\textbf{\textit{Q}}'} \ {\big\langle} c^\dag_{\zeta_1,\textbf{\textit{K}}_1+\alpha_{x\text{-}e}\textbf{\textit{k}}_2+\alpha_{x}\textbf{\textit{Q}}_2+\textbf{\textit{Q}}'} v^{\phantom{\dagger}}_{\zeta_2,\textbf{\textit{K}}_1-\beta_{x\text{-}e}\textbf{\textit{k}}_2-\beta_{x}\textbf{\textit{Q}}_2} c^\dag_{\zeta_3,\alpha_{x\text{-}e}\textbf{\textit{k}}_2-\textbf{\textit{Q}}_2-\textbf{\textit{Q}}'} c^{\phantom{\dagger}}_{\zeta_4,\textbf{\textit{k}}_2} {\big\rangle}^{c} \notag \\
			& & \hspace{3.9mm} - \frac{i}{2\hbar} \sum_{\textbf{\textit{Q}}'} W_{\textbf{\textit{Q}}'} \ {\big\langle} c^\dag_{\zeta_1,\textbf{\textit{K}}_1+\alpha_{x\text{-}e}\textbf{\textit{k}}_2+\alpha_{x}\textbf{\textit{Q}}_2+\textbf{\textit{Q}}'} v^{\phantom{\dagger}}_{\zeta_2,\textbf{\textit{K}}_1-\beta_{x\text{-}e}\textbf{\textit{k}}_2-\beta_{x}\textbf{\textit{Q}}_2+\textbf{\textit{Q}}'} c^\dag_{\zeta_3,\alpha_{x\text{-}e}\textbf{\textit{k}}_2-\textbf{\textit{Q}}_2} c^{\phantom{\dagger}}_{\zeta_4,\textbf{\textit{k}}_2} {\big\rangle}^{c} \notag \\
			& & \hspace{3.9mm} - \frac{i}{2\hbar} \sum_{\textbf{\textit{Q}}'} W_{\textbf{\textit{Q}}'} \ {\big\langle} c^\dag_{\zeta_1,\textbf{\textit{K}}_1+\alpha_{x\text{-}e}\textbf{\textit{k}}_2+\alpha_{x}\textbf{\textit{Q}}_2} v^{\phantom{\dagger}}_{\zeta_2,\textbf{\textit{K}}_1-\beta_{x\text{-}e}\textbf{\textit{k}}_2-\beta_{x}\textbf{\textit{Q}}_2+\textbf{\textit{Q}}'} c^\dag_{\zeta_3,\alpha_{x\text{-}e}\textbf{\textit{k}}_2-\textbf{\textit{Q}}_2+\textbf{\textit{Q}}'} c^{\phantom{\dagger}}_{\zeta_4,\textbf{\textit{k}}_2} {\big\rangle}^{c} . \label{eq:ass-tran-3}
			\end{eqnarray}
			Next, the dynamics of symmetric ``$+$'' and antisymmetric ``$-$'' linear combinations of Eq.~\eqref{eq:ass-tran-3} is considered:
			\begin{eqnarray}
			& & \left\lbrace\partial_t - \frac{i}{\hbar} \left[ \varepsilon\,^{\zeta_1}_g + \frac{\hbar^2\textbf{\textit{K}}_1^2}{2\mu} -\frac{\hbar^2\textbf{\textit{k}}_2^2}{2}\frac{m_e+m_h}{(2m_e+m_h)m_e} +\frac{\hbar^2\textbf{\textit{Q}}_2^2}{2}\left(\frac{1}{m_e+m_h}+\frac{1}{m_e}\right) \right] \right\rbrace \notag \\
			& & \times \frac{1}{2} \Big({\big\langle} c^\dag_{\zeta_1,\textbf{\textit{K}}_1+\alpha_{x\text{-}e}\textbf{\textit{k}}_2+\alpha_{x}\textbf{\textit{Q}}_2} v^{\phantom{\dagger}}_{\zeta_1,\textbf{\textit{K}}_1-\beta_{x\text{-}e}\textbf{\textit{k}}_2-\beta_{x}\textbf{\textit{Q}}_2} c^\dag_{\zeta_2,\alpha_{x\text{-}e}\textbf{\textit{k}}_2-\textbf{\textit{Q}}_2} c^{\phantom{\dagger}}_{\zeta_2,\textbf{\textit{k}}_2} {\big\rangle}^{c} \notag \\ & & \hspace{7.6mm} \pm {\big\langle} c^\dag_{\zeta_2,\textbf{\textit{K}}_1+\alpha_{x\text{-}e}\textbf{\textit{k}}_2+\alpha_{x}\textbf{\textit{Q}}_2} v^{\phantom{\dagger}}_{\zeta_1,\textbf{\textit{K}}_1-\beta_{x\text{-}e}\textbf{\textit{k}}_2-\beta_{x}\textbf{\textit{Q}}_2} c^\dag_{\zeta_1,\alpha_{x\text{-}e}\textbf{\textit{k}}_2-\textbf{\textit{Q}}_2} c^{\phantom{\dagger}}_{\zeta_2,\textbf{\textit{k}}_2} {\big\rangle}^{c}\Big) \notag \\
			& & = \frac{i\left(1\pm\delta_{\zeta_1,\zeta_2}\right)}{2\hbar} \ W_{(1-\alpha_{x\text{-}e})\textbf{\textit{k}}_2+\textbf{\textit{Q}}_2} \Big({\big\langle} c^\dag_{\zeta_1,\textbf{\textit{K}}_1-\beta_{x\text{-}e}\textbf{\textit{k}}_2-\beta_{x}\textbf{\textit{Q}}_2} v^{\phantom{\dagger}}_{\zeta_1,\textbf{\textit{K}}_1-\beta_{x\text{-}e}\textbf{\textit{k}}_2-\beta_{x}\textbf{\textit{Q}}_2} {\big\rangle} \notag \\ & & \hspace{49.8mm} - {\big\langle} c^\dag_{\zeta_1,\textbf{\textit{K}}_1+\alpha_{x\text{-}e}\textbf{\textit{k}}_2+\alpha_{x}\textbf{\textit{Q}}_2} v^{\phantom{\dagger}}_{\zeta_1,\textbf{\textit{K}}_1+\alpha_{x\text{-}e}\textbf{\textit{k}}_2+\alpha_{x}\textbf{\textit{Q}}_2} {\big\rangle}\Big)f\,^{\zeta_2}_{e,\textbf{\textit{k}}_2} \notag \\
			& & \hspace{3.9mm} \mp \frac{i\left(1\pm\delta_{\zeta_1,\zeta_2}\right)}{2\hbar} \ W_{\textbf{\textit{K}}_1-(1-\alpha_{x\text{-}e})\textbf{\textit{k}}_2+\alpha_{x}\textbf{\textit{Q}}_2} \Big({\big\langle} c^\dag_{\zeta_1,\textbf{\textit{K}}_1-\beta_{x\text{-}e}\textbf{\textit{k}}_2-\beta_{x}\textbf{\textit{Q}}_2} v^{\phantom{\dagger}}_{\zeta_1,\textbf{\textit{K}}_1-\beta_{x\text{-}e}\textbf{\textit{k}}_2-\beta_{x}\textbf{\textit{Q}}_2} {\big\rangle} \notag \\ & & \hspace{62.2mm} - {\big\langle} c^\dag_{\zeta_1,\alpha_{x\text{-}e}\textbf{\textit{k}}_2-\textbf{\textit{Q}}_2} v^{\phantom{\dagger}}_{\zeta_1,\alpha_{x\text{-}e}\textbf{\textit{k}}_2-\textbf{\textit{Q}}_2} {\big\rangle}\Big)f\,^{\zeta_2}_{e,\textbf{\textit{k}}_2} \notag \\
			& & \hspace{3.9mm} + \frac{i}{2\hbar} \sum_{\textbf{\textit{Q}}'} W_{\textbf{\textit{Q}}'} \Big({\big\langle} c^\dag_{\zeta_1,\textbf{\textit{K}}_1+\alpha_{x\text{-}e}\textbf{\textit{k}}_2+\alpha_{x}\textbf{\textit{Q}}_2+\textbf{\textit{Q}}'} v^{\phantom{\dagger}}_{\zeta_1,\textbf{\textit{K}}_1-\beta_{x\text{-}e}\textbf{\textit{k}}_2-\beta_{x}\textbf{\textit{Q}}_2} c^\dag_{\zeta_2,\alpha_{x\text{-}e}\textbf{\textit{k}}_2-\textbf{\textit{Q}}_2-\textbf{\textit{Q}}'} c^{\phantom{\dagger}}_{\zeta_2,\textbf{\textit{k}}_2} {\big\rangle}^{c} \notag \\
			& & \hspace{26.4mm} \pm {\big\langle} c^\dag_{\zeta_2,\textbf{\textit{K}}_1+\alpha_{x\text{-}e}\textbf{\textit{k}}_2+\alpha_{x}\textbf{\textit{Q}}_2+\textbf{\textit{Q}}'} v^{\phantom{\dagger}}_{\zeta_1,\textbf{\textit{K}}_1-\beta_{x\text{-}e}\textbf{\textit{k}}_2-\beta_{x}\textbf{\textit{Q}}_2} c^\dag_{\zeta_1,\alpha_{x\text{-}e}\textbf{\textit{k}}_2-\textbf{\textit{Q}}_2-\textbf{\textit{Q}}'} c^{\phantom{\dagger}}_{\zeta_2,\textbf{\textit{k}}_2} {\big\rangle}^{c}\Big) \notag \\
			& & \hspace{3.9mm} - \frac{i}{2\hbar} \sum_{\textbf{\textit{Q}}'} W_{\textbf{\textit{Q}}'} \Big({\big\langle} c^\dag_{\zeta_1,\textbf{\textit{K}}_1+\alpha_{x\text{-}e}\textbf{\textit{k}}_2+\alpha_{x}\textbf{\textit{Q}}_2+\textbf{\textit{Q}}'} v^{\phantom{\dagger}}_{\zeta_1,\textbf{\textit{K}}_1-\beta_{x\text{-}e}\textbf{\textit{k}}_2-\beta_{x}\textbf{\textit{Q}}_2+\textbf{\textit{Q}}'} c^\dag_{\zeta_2,\alpha_{x\text{-}e}\textbf{\textit{k}}_2-\textbf{\textit{Q}}_2} c^{\phantom{\dagger}}_{\zeta_2,\textbf{\textit{k}}_2} {\big\rangle}^{c} \notag \\
			& & \hspace{26.4mm} \pm {\big\langle} c^\dag_{\zeta_2,\textbf{\textit{K}}_1+\alpha_{x\text{-}e}\textbf{\textit{k}}_2+\alpha_{x}\textbf{\textit{Q}}_2+\textbf{\textit{Q}}'} v^{\phantom{\dagger}}_{\zeta_1,\textbf{\textit{K}}_1-\beta_{x\text{-}e}\textbf{\textit{k}}_2-\beta_{x}\textbf{\textit{Q}}_2+\textbf{\textit{Q}}'} c^\dag_{\zeta_1,\alpha_{x\text{-}e}\textbf{\textit{k}}_2-\textbf{\textit{Q}}_2} c^{\phantom{\dagger}}_{\zeta_2,\textbf{\textit{k}}_2} {\big\rangle}^{c}\Big) \notag \\
			& & \hspace{3.9mm} - \frac{i}{2\hbar} \sum_{\textbf{\textit{Q}}'} W_{\textbf{\textit{Q}}'} \Big({\big\langle} c^\dag_{\zeta_1,\textbf{\textit{K}}_1+\alpha_{x\text{-}e}\textbf{\textit{k}}_2+\alpha_{x}\textbf{\textit{Q}}_2} v^{\phantom{\dagger}}_{\zeta_1,\textbf{\textit{K}}_1-\beta_{x\text{-}e}\textbf{\textit{k}}_2-\beta_{x}\textbf{\textit{Q}}_2+\textbf{\textit{Q}}'} c^\dag_{\zeta_2,\alpha_{x\text{-}e}\textbf{\textit{k}}_2-\textbf{\textit{Q}}_2+\textbf{\textit{Q}}'} c^{\phantom{\dagger}}_{\zeta_2,\textbf{\textit{k}}_2} {\big\rangle}^{c} \notag \\
			& & \hspace{26.4mm} \pm {\big\langle} c^\dag_{\zeta_2,\textbf{\textit{K}}_1+\alpha_{x\text{-}e}\textbf{\textit{k}}_2+\alpha_{x}\textbf{\textit{Q}}_2} v^{\phantom{\dagger}}_{\zeta_1,\textbf{\textit{K}}_1-\beta_{x\text{-}e}\textbf{\textit{k}}_2-\beta_{x}\textbf{\textit{Q}}_2+\textbf{\textit{Q}}'} c^\dag_{\zeta_1,\alpha_{x\text{-}e}\textbf{\textit{k}}_2-\textbf{\textit{Q}}_2+\textbf{\textit{Q}}'} c^{\phantom{\dagger}}_{\zeta_2,\textbf{\textit{k}}_2} {\big\rangle}^{c}\Big) .  \label{eq:ass-tran-4}
			\end{eqnarray}
			The interband transitions on the right-hand side of Eq.~\eqref{eq:ass-tran-4} are now expanded in terms of the complete set of exciton wave functions and exciton transitions following Eq.~\eqref{eq:definition-exciton}.			
			Furthermore, the symmetric ``$+$'' and antisymmetric ``$-$'' linear combinations of electron-density-assisted transitions in Eq.~\eqref{eq:ass-tran-4} are expressed by exciton wave functions $\varphi^R\,^{\zeta_1}_{\nu_3,\textbf{\textit{K}}_1}$, expansion coefficients $\hat{T}\,^{\zeta_1,\zeta_2}_{x\text{-}e,\pm,\nu_3,\textbf{\textit{Q}}_2}$, and electron densities $f\,^{\zeta_2}_{e,\textbf{\textit{k}}_2}$ according to Eq.~\eqref{eq:lin-comb}:
			\begin{eqnarray}
			& & \sum_{\nu_3} \bigg\lbrace\partial_t - \frac{i}{\hbar} \bigg[\epsilon_{x,\zeta_1,\nu_3} -\frac{\hbar^2\textbf{\textit{k}}_2^2}{2}\frac{m_e+m_h}{(2m_e+m_h)m_e} +\frac{\hbar^2\textbf{\textit{Q}}_2^2}{2}\bigg(\frac{1}{m_e+m_h}+\frac{1}{m_e}\bigg)\bigg] \bigg\rbrace \varphi^R\,^{\zeta_1}_{\nu_3,\textbf{\textit{K}}_1} \ f\,^{\zeta_2}_{e,\textbf{\textit{k}}_2} \ \hat{T}\,^{\zeta_1,\zeta_2}_{x\text{-}e,\pm,\nu_3,\textbf{\textit{Q}}_2} \notag \\
			& & \mp \sum_{\nu_3} \bigg\lbrace\partial_t - \frac{i}{\hbar} \bigg[\epsilon_{x,\zeta_1,\nu_3} -\frac{\hbar^2\textbf{\textit{k}}_2^2}{2}\frac{m_e+m_h}{(2m_e+m_h)m_e} +\frac{\hbar^2\left(-\textbf{\textit{K}}_1-\alpha_{x}\textbf{\textit{Q}}_2\right)^2}{2}\bigg(\frac{1}{m_e+m_h}+\frac{1}{m_e}\bigg)\bigg] \bigg\rbrace \notag \\
			& & \hspace{9mm} \times \varphi^R\,^{\zeta_1}_{\nu_3,\alpha_{x}\textbf{\textit{K}}_1+[\alpha_{x}^2-1]\textbf{\textit{Q}}_2} \ f\,^{\zeta_2}_{e,\textbf{\textit{k}}_2}  \ \hat{T}\,^{\zeta_1,\zeta_2}_{x\text{-}e,\pm,\nu_3,-\textbf{\textit{K}}_1-\alpha_{x}\textbf{\textit{Q}}_2} \notag \\
			& & = \frac{i\left(1\pm\delta_{\zeta_1,\zeta_2}\right)}{2\hbar} \sum_{\nu_3} \Big[ W_{(1-\alpha_{x\text{-}e})\textbf{\textit{k}}_2+\textbf{\textit{Q}}_2} \left(\varphi^R\,^{\zeta_1}_{\nu_3,\textbf{\textit{K}}_1-\beta_{x\text{-}e}\textbf{\textit{k}}_2-\beta_{x}\textbf{\textit{Q}}_2} - \varphi^R\,^{\zeta_1}_{\nu_3,\textbf{\textit{K}}_1+\alpha_{x\text{-}e}\textbf{\textit{k}}_2+\alpha_{x}\textbf{\textit{Q}}_2}\right) \notag \\
			& & \hspace{31.1mm} \mp W_{\textbf{\textit{K}}_1-(1-\alpha_{x\text{-}e})\textbf{\textit{k}}_2+\alpha_{x}\textbf{\textit{Q}}_2} \left(\varphi^R\,^{\zeta_1}_{\nu_3,\textbf{\textit{K}}_1-\beta_{x\text{-}e}\textbf{\textit{k}}_2-\beta_{x}\textbf{\textit{Q}}_2} - \varphi^R\,^{\zeta_1}_{\nu_3,\alpha_{x\text{-}e}\textbf{\textit{k}}_2-\textbf{\textit{Q}}_2}\right) \Big] f\,^{\zeta_2}_{e,\textbf{\textit{k}}_2} \ P\,^{\zeta_1}_{\nu_3} \notag \\
			& & \hspace{3.9mm} + \frac{i}{\hbar} \sum_{\nu_3,\textbf{\textit{Q}}_3} \Big[ W_{\textbf{\textit{Q}}_2-\textbf{\textit{Q}}_3} \left(\varphi^R\,^{\zeta_1}_{\nu_3,\textbf{\textit{K}}_1-\beta_{x}(\textbf{\textit{Q}}_2-\textbf{\textit{Q}}_3)} - \varphi^R\,^{\zeta_1}_{\nu_3,\textbf{\textit{K}}_1+\alpha_{x}(\textbf{\textit{Q}}_2-\textbf{\textit{Q}}_3)}\right) \notag \\
			& & \hspace{19.5mm} \mp W_{\textbf{\textit{K}}_1+\alpha_{x}\textbf{\textit{Q}}_2+\textbf{\textit{Q}}_3} \left(\varphi^R\,^{\zeta_1}_{\nu_3,\textbf{\textit{K}}_1-\beta_{x}(\textbf{\textit{Q}}_2-\textbf{\textit{Q}}_3)} - \varphi^R\,^{\zeta_1}_{\nu_3,-\textbf{\textit{Q}}_2-\alpha_{x}\textbf{\textit{Q}}_3}\right) \Big] f\,^{\zeta_2}_{e,\textbf{\textit{k}}_2} \ \hat{T}\,^{\zeta_1,\zeta_2}_{x\text{-}e,\pm,\nu_3,\textbf{\textit{Q}}_3} . \label{eq:ass-tran-5}
			\end{eqnarray}
			Here, we used the following relation which directly follows from the definitions of the reduced mass $\mu = m_{e}m_{h}/(m_{e}+m_{h})$ and the ratio $\alpha_{x}$ given in Eq.~\eqref{eq:alpha-beta}:
			\begin{equation}
			\frac{\textbf{\textit{K}}_1^2}{\mu} +\textbf{\textit{Q}}_2^2\left(\frac{1}{m_e+m_h}+\frac{1}{m_e}\right)
			= \frac{(\alpha_{x}\textbf{\textit{K}}_1+[\alpha_{x}^2-1]\textbf{\textit{Q}}_2)^2}{\mu} +(-\textbf{\textit{K}}_1-\alpha_{x}\textbf{\textit{Q}}_2)^2\left(\frac{1}{m_e+m_h}+\frac{1}{m_e}\right) .
			\end{equation}
			Note that the new basis defined by Eq.~\eqref{eq:lin-comb} now implicitly includes two Coulomb contributions which revealed the structure of the Wannier equation, Eq.~\eqref{eq:Wannier-Gl}.
			Therefore, the last two lines of Eq.~\eqref{eq:ass-tran-5} only encompass four contributions instead of previously six electron-density-assisted transitions characterized by the last six lines of Eq.~\eqref{eq:ass-tran-4}.
			Next, Eq.~\eqref{eq:ass-tran-5} is multiplied by $\frac{1}{\mathcal{A}\sum_{\textbf{\textit{k}}_3} f\,^{\zeta_2}_{e,\textbf{\textit{k}}_3}} \sum_{\textbf{\textit{K}}_1}\varphi^L\,^{\zeta_1}_{\nu_2,\textbf{\textit{K}}_1} \sum_{\textbf{\textit{k}}_2}$ and the normalization of exciton wave functions, Eq.~\eqref{eq:normierung-x}, is used to obtain:
			\begin{eqnarray}
			& & \sum_{\nu_3,\textbf{\textit{K}}_1} S\,^{\zeta_1}_{x\text{-}e,\pm,\nu_2,\nu_3,\textbf{\textit{Q}}_2,\textbf{\textit{K}}_1} \left\lbrace\partial_t - \frac{i}{\hbar} \left[\epsilon_{x,\zeta_1,\nu_3} +\frac{\hbar^2\textbf{\textit{K}}_1^2}{2}\left(\frac{1}{m_e+m_h}+\frac{1}{m_e}\right) -\Delta\,^{\zeta_2}_{e}\right]\right\rbrace \hat{T}\,^{\zeta_1,\zeta_2}_{x\text{-}e,\pm,\nu_3,\textbf{\textit{K}}_1} \notag \\
			& & = \frac{i\left(1\pm\delta_{\zeta_1,\zeta_2}\right)}{2\hbar\mathcal{A}\sum_{\textbf{\textit{k}}_1} f\,^{\zeta_2}_{e,\textbf{\textit{k}}_1}} \sum_{\nu_3,\textbf{\textit{k}}_2} \left(\hat{W}\,^{\zeta_1}_{x\text{-}e,\pm,\nu_2,\nu_3,\textbf{\textit{Q}}_2,\textbf{\textit{k}}_2}\right)^*f\,^{\zeta_2}_{e,\textbf{\textit{k}}_2} \ P\,^{\zeta_1}_{\nu_3}
			+ \frac{i}{\hbar}  \sum_{\nu_3,\textbf{\textit{Q}}_3} W\,^{\zeta_1}_{x\text{-}e,\pm,\nu_2,\nu_3,\textbf{\textit{Q}}_2,\textbf{\textit{Q}}_3} \ \hat{T}\,^{\zeta_1,\zeta_2}_{x\text{-}e,\pm,\nu_3,\textbf{\textit{Q}}_3} . \label{eq:zw-eq-5}
			\end{eqnarray}
			The definitions of $S\,^{\zeta_1}_{x\text{-}e,\pm,\nu_2,\nu_3,\textbf{\textit{Q}}_2,\textbf{\textit{K}}_1}$, $\Delta\,^{\zeta_2}_{e}$, $\hat{W}\,^{\zeta_1}_{x\text{-}e,\pm,\nu_2,\nu_3,\textbf{\textit{Q}}_2,\textbf{\textit{k}}_2}$, and $W\,^{\zeta_1}_{x\text{-}e,\pm,\nu_2,\nu_3,\textbf{\textit{Q}}_2,\textbf{\textit{Q}}_3}$ are given in Eqs.~\eqref{eq:S-Matrix}, \eqref{eq:e-renorm}, \eqref{eq:mat-el-x-e}, and \eqref{eq:Coulomb-M-E}, respectively.
			Finally, Eq.~\eqref{eq:zw-eq-5} is multiplied by $\sum_{\nu_2,\textbf{\textit{Q}}_2} \big(S\,^{\zeta_1}_{x\text{-}e,\pm}\big)^{-1}_{\nu_1,\nu_2,\textbf{\textit{Q}}_1,\textbf{\textit{Q}}_2}$ and one obtains:
			\begin{eqnarray}
			& & \left\lbrace \partial_t - \frac{i}{\hbar}\left[\epsilon_{x,\zeta_1,\nu_1} +\frac{\hbar^2\textbf{\textit{Q}}_1^2}{2}\left(\frac{1}{m_e+m_h}+\frac{1}{m_e}\right) -\Delta\,^{\zeta_2}_{e}\right] \right\rbrace \hat{T}\,^{\zeta_1,\zeta_2}_{x\text{-}e,\pm,\nu_1,\textbf{\textit{Q}}_1} \notag \\
			& &= \frac{i\left(1\pm\delta_{\zeta_1,\zeta_2}\right)}{2\hbar\mathcal{A}\sum_{\textbf{\textit{k}}_1} f\,^{\zeta_2}_{e,\textbf{\textit{k}}_1}} \sum_{\nu_2,\textbf{\textit{Q}}_2} \big(S\,^{\zeta_1}_{x\text{-}e,\pm}\big)^{-1}_{\nu_1,\nu_2,\textbf{\textit{Q}}_1,\textbf{\textit{Q}}_2} \sum_{\nu_3,\textbf{\textit{k}}_2} \left(\hat{W}\,^{\zeta_1}_{x\text{-}e,\pm,\nu_2,\nu_3,\textbf{\textit{Q}}_2,\textbf{\textit{k}}_2}\right)^*f\,^{\zeta_2}_{e,\textbf{\textit{k}}_2} \ P\,^{\zeta_1}_{\nu_3} \notag \\
			& & \hspace{3.9mm} + \frac{i}{\hbar} \sum_{\nu_2,\textbf{\textit{Q}}_2} \big(S\,^{\zeta_1}_{x\text{-}e,\pm}\big)^{-1}_{\nu_1,\nu_2,\textbf{\textit{Q}}_1,\textbf{\textit{Q}}_2} \sum_{\nu_3,\textbf{\textit{Q}}_3} W\,^{\zeta_1}_{x\text{-}e,\pm,\nu_2,\nu_3,\textbf{\textit{Q}}_2,\textbf{\textit{Q}}_3} \ \hat{T}\,^{\zeta_1,\zeta_2}_{x\text{-}e,\pm,\nu_3,\textbf{\textit{Q}}_3} .\label{eq:zw-eq-6}
			\end{eqnarray}
			The inverse of the matrix $S\,^{\zeta_1}_{x\text{-}e,\pm,\nu_1,\nu_2,\textbf{\textit{Q}}_1,\textbf{\textit{Q}}_2}$ is defined by:
			\begin{equation}
			\sum_{\nu_2,\textbf{\textit{Q}}_2} \big(S\,^{\zeta_1}_{x\text{-}e,\pm}\big)^{-1}_{\nu_1,\nu_2,\textbf{\textit{Q}}_1,\textbf{\textit{Q}}_2} \ S\,^{\zeta_1}_{x\text{-}e,\pm,\nu_2,\nu_3,\textbf{\textit{Q}}_2,\textbf{\textit{Q}}_3} = \delta_{\nu_1,\nu_3} \ \delta_{\textbf{\textit{Q}}_1,\textbf{\textit{Q}}_3} .
			\end{equation}
			Due to the conveniently chosen coordinates as well as symmetric ``$+$'' and antisymmetric ``$-$'' linear combinations, the Schrödinger equation given in Eq.~\eqref{eq:Bi-X} can be identified in Eq.~\eqref{eq:zw-eq-6}.
			This Schrödinger equation, Eq.~\eqref{eq:Bi-X}, only depends on the wave vector $\textbf{\textit{Q}}_1$ and can be directly solved.
			Expanding the coefficients $\hat{T}\,^{\zeta_1,\zeta_2}_{x\text{-}e,\pm,\nu_1,\textbf{\textit{Q}}}$ in terms of corresponding wave functions $\psi^R\,^{\zeta_1,\zeta_2}_{x\text{-}e,\pm,\mu,\nu_1,\textbf{\textit{Q}}}$ and new expansion coefficients $T\,^{\zeta_1,\zeta_2}_{x\text{-}e,\pm,\mu}$ according to Eq.~\eqref{eq:expansion} leads to Eq.~\eqref{eq:neg-trion-dyn}.	

\section{Angular Momentum} \label{subsec:magn-quant}

	Neglecting band structure asymmetries of the atomically thin semiconductor in the in-plane directions around the high-symmetry points \cite{kormanyos2013monolayer} leads to an in-plane rotational symmetry of the Wannier equation, Eq.~\eqref{eq:Wannier-Gl}.
	Consequently, the exciton wave functions $\varphi^R\,^{\zeta_1}_{\nu,\textbf{\textit{k}}}$ can be separated into the product of a radial part $\Phi^R\,^{\zeta_1}_{\nu,k}$ and a phase factor $e^{i m_{\nu} \phi_k}$:
	\begin{equation}
	\varphi^R\,^{\zeta_1}_{\nu,\textbf{\textit{k}}} = \Phi^R\,^{\zeta_1}_{\nu,k} \ e^{i m_{\nu} \phi_k} .
	\end{equation}
	The radial $k$ and angular coordinates $\phi_k$ are defined by $\textbf{\textit{k}} = k \, e^{i \phi_k}$.
	The phase $e^{i m_{\nu} \phi_k}$ is determined by the angular momentum $m_{\nu}$ associated with the exciton eigenstate $\nu$.
	On the other hand, the trion and exciton-electron/hole continuum wave functions $\psi^R\,^{\zeta_1,\zeta_2}_{x\text{-}e/h,\pm,\mu,\nu,\textbf{\textit{k}}}$ are expanded in a Fourier series in angle coordinates \cite{takayama2002t,schumacher2006coherent} as follows:
	\begin{equation}
	\psi^R\,^{\zeta_1,\zeta_2}_{x\text{-}e/h,\pm,\mu,\nu,\textbf{\textit{k}}} = \sum_{m_{\mu}} \Psi^R\,^{\zeta_1,\zeta_2}_{x\text{-}e/h,\pm,\mu,\nu,k} \ e^{i m_{\mu} \phi_k} ,
	\end{equation}
	with the angular momentum $m_{\mu}$.
	As a result, the matrix elements in the eigenvalue Eq.~\eqref{eq:Bi-X} are also expanded in a Fourier series which obey angular momentum conservation which we have also verified numerically.
	Our numerical evaluations are restricted to the energetically lowest $\nu = 1s$, $2s$, and $2p^\pm$ exciton states with angular momentum $m_{\nu}= 0$, $\pm1$ and associated trion states with angular momentum $m_{\mu}= 0$, $\pm1$.

\section{Analytical Solution in Frequency Domain} \label{app:fourier-transform}

	The set of coupled differential equations, Eq.~\eqref{eq:chi-3-pol} and \eqref{eq:neg-trion-dyn}, can be analytically solved in frequency domain.
	The exciton transitions $P\,^{\zeta_1}_{\nu_1}$, described by Eq.~\eqref{eq:chi-3-pol}, are optically driven by the light field at the monolayer position ${{E}}^{\sigma_j}_{T}(t)$ which enters the equation of motion via the Rabi frequency $\Omega\,^{\zeta_1,\sigma_j}_{\nu_1,\textbf{\textit{k}}}$ defined in Eq.~\eqref{eq:Rabi-freq}.
	The light field ${{E}}^{\sigma_j}_{T}(t)$ is determined by the incoming light field ${E}_0^{\sigma_j} (t)$ and the interband polarization $P\,^{\sigma_j}_{} \left(t\right)$ \cite{knorr1996theory,jahnke1997linear}.
	After transferring the light fields ${{E}}^{\sigma_j}_{0/T}(t) = \tilde{{E}}^{\sigma_j}_{0/T}(t) \ e^{-i\omega_0 t}$ and the interband polarization $P\,^{\sigma_j}_{} \left(t\right) = \tilde{P}\,^{\sigma_j}_{} \left(t\right) \ e^{-i\omega_0 t}$ into a rotating frame of the laser frequency $\omega_0$ and applying a slowly varying envelope approximation \cite{stroucken1996coherent}, the envelope of the light field $\tilde{{E}}^{\sigma_j}_{T}(t)$ becomes:
	\begin{equation}
	\tilde{{E}}^{\sigma_j}_{T}(t)
	= \tilde{E}_0^{\sigma_j} (t)
	+i  \frac{\omega_0}{2\varepsilon_0c_0n_r\mathcal{A}} \sum_{\zeta_1,\nu_1,\textbf{\textit{k}}_1} \varphi^L\,^{\zeta_1}_{\nu_1,\textbf{\textit{k}}_1} \left[ d\,^{\zeta_1,\sigma_j}_{c,v\vphantom{\textbf{\textit{k}}_1}} \, \tilde{P}\,^{\zeta_1}_{\nu_1}\left(t\right)\right]^* \label{eq:trans} .
	\end{equation}
	Here, $n_r$ is the constant background refractive index, the macroscopic interband polarization $P\,^{\sigma_j}_{} \left(t\right)$ was already expanded according to Eq.~\eqref{eq:makr-interband-pol1}, and counter-rotating terms were neglected \cite{scully1997quantum}.
	Equation~\eqref{eq:trans} is subsequently inserted into the Rabi frequency $\Omega\,^{\zeta_1,\sigma_j}_{\nu_1,\textbf{\textit{k}}}$ given in Eq.~\eqref{eq:Rabi-freq}.
	Afterwards, the incoming light field $ \hat{E}_{0}^{\sigma_j}(\omega) = \int dt \ e^{i(\omega-\omega_0)t} \ \tilde{E}_0^{\sigma_j}(t)$, exciton transitions $\hat{P}^*\,^{\zeta_1}_{\nu_1}(\omega) = \int dt \ e^{i(\omega-\omega_0)t} \ \tilde{P}^*\,^{\zeta_1}_{\nu_1}(t)$, and trion and exciton-electron/hole continuum states $\hat{T}^*\,^{\zeta_1,\zeta_2}_{x\text{-}e/h,\pm,\mu}(\omega) = \int dt \ e^{i(\omega-\omega_0)t} \ \tilde{T}^*\,^{\zeta_1,\zeta_2}_{x\text{-}e/h,\pm,\mu}(t)$ are Fourier transformed.
	Consequently, the result of solving the trion and exciton-electron/hole continuum dynamics, described by Eq.~\eqref{eq:neg-trion-dyn}, is inserted into the equation of motion of the exciton transitions, determined by Eq.~\eqref{eq:chi-3-pol}.
	The procedure leads to the exciton transitions $\hat{P}^*\,^{\zeta_1}_{\nu_1}(\omega)$ in frequency domain:
	\begin{equation}
	\left[\Gamma\,^{\zeta_1}_{x} + i \left(\epsilon\,^{\zeta_1}_{x,\nu_1}-\hbar\omega\right) \right] \hat{P}^*\,^{\zeta_1}_{\nu_1}(\omega)
	= i \hat{d}\,^{\zeta_1,\sigma_j}_{c,v,\nu_1} \ \hat{E}_{0}^{\sigma_j}(\omega)
	 - \sum_{\nu_2} \Big[ \Gamma\,^{\zeta_1}_{r\text{-}e/h,\nu_1,\nu_2} + \Gamma\,^{\zeta_1}_{x\text{-}e/h,\nu_1,\nu_2}(\omega) + i\Delta\,^{\zeta_1}_{x\text{-}e/h,\nu_1,\nu_2}(\omega) \Big] \hat{P}^*\,^{\zeta_1}_{\nu_2}(\omega) , \label{eq:exc-freq}
	\end{equation}
	with the phonon-mediated dephasing $\Gamma\,^{\zeta_1}_{x} = \hbar\gamma\,^{\zeta_1}_{x}$ and the dipole transition element $\hat{d}\,^{\zeta_1,\sigma_j}_{c,v,\nu_1}$:
	\begin{equation}
	\hat{d}\,^{\zeta_1,\sigma_j}_{c,v,\nu_1} = \frac{1}{\mathcal{A}} \sum_{\textbf{\textit{k}}} \varphi^R\,^{\zeta_1}_{\nu_1,\textbf{\textit{k}}} \ d\,^{\zeta_1,\sigma_j}_{c,v\vphantom{\textbf{\textit{k}}_1}} \big(1-f\,^{\zeta_1}_{e/h,\textbf{\textit{k}}}\big) .
	\end{equation}
	Equation~\eqref{eq:exc-freq} couples exciton transitions $\hat{P}^*\,^{\zeta_1}_{\nu_1}(\omega)$ with different quantum numbers $\nu_1$ and $\nu_2$ via the radiative dephasing $\Gamma\,^{\zeta_1}_{r\text{-}e/h,\nu_1,\nu_2}$, the frequency-dependent scattering-induced dephasing $\Gamma\,^{\zeta_1}_{x\text{-}e/h,\nu_1,\nu_2}(\omega)$ as well as the frequency-dependent renormalization $\Delta\,^{\zeta_1}_{x\text{-}e/h,\nu_1,\nu_2}(\omega)$ defined by:
	\begin{eqnarray}
	\Gamma\,^{\zeta_1}_{r\text{-}e/h,\nu_1,\nu_2} & = & \frac{\omega_0 \big|d\,^{\zeta_1,\sigma_j}_{c,v\vphantom{\textbf{\textit{k}}_1}}\big|^2}{2\varepsilon_0c_0n_r\mathcal{A}^2} \sum_{\textbf{\textit{k}}_1} \varphi^R\,^{\zeta_1}_{\nu_1,\textbf{\textit{k}}_1} \big(1-f\,^{\zeta_1}_{e/h,\textbf{\textit{k}}_1}\big) \sum_{\nu_2,\textbf{\textit{k}}_2} \varphi^L\,^{\zeta_1}_{\nu_2,\textbf{\textit{k}}_2} , \\
	\Gamma\,^{\zeta_1}_{x\text{-}e/h,\nu_1,\nu_2}(\omega)
	& = & \frac{\pi}{\mathcal{A}} \sum_{\zeta_2,\pm,\mu} \mathcal{W}\,^{\zeta_1,\zeta_2}_{x\text{-}e/h,\pm,\mu,\nu_1} \ \hat{\mathcal{W}}\,^{\zeta_1,\zeta_2}_{x\text{-}e/h,\pm,\mu,\nu_2} \ \delta_{\hbar\gamma\,^{\zeta_1}_{x\text{-}e/h}} \big(\epsilon\,^{\zeta_1,\zeta_2}_{x\text{-}e/h,\pm,\mu} + \Delta\,^{\zeta_2}_{e/h}-\hbar\omega\big) , \\
	\Delta\,^{\zeta_1}_{x\text{-}e/h,\nu_1,\nu_2}(\omega)
	& = & \frac{1}{\mathcal{A}} \sum_{\textbf{\textit{k}}} \hat{W}\,^{\zeta_1}_{H\text{-}F,\nu_2,\nu_1,\textbf{\textit{k}}} \ f\,^{\zeta_1}_{e/h,\textbf{\textit{k}}} - \frac{1}{\mathcal{A}}\sum_{\zeta_2,\pm,\mu}  \frac{\mathcal{W}\,^{\zeta_1,\zeta_2}_{x\text{-}e/h,\pm,\mu,\nu_1} \ \hat{\mathcal{W}}\,^{\zeta_1,\zeta_2}_{x\text{-}e/h,\pm,\mu,\nu_2} \big(\epsilon\,^{\zeta_1,\zeta_2}_{x\text{-}e/h,\pm,\mu} + \Delta\,^{\zeta_2}_{e/h}-\hbar\omega\big)}{\big(\hbar\gamma\,^{\zeta_1}_{x\text{-}e/h}\big)^2 + \big(\epsilon\,^{\zeta_1,\zeta_2}_{x\text{-}e/h,\pm,\mu} + \Delta\,^{\zeta_2}_{e/h}-\hbar\omega\big)^2} . \arxivcustomneu
	\end{eqnarray}
	The abbreviations $\mathcal{W}\,^{\zeta_1,\zeta_2}_{x\text{-}e/h,\pm,\mu,\nu_1}$ and $\hat{\mathcal{W}}\,^{\zeta_1,\zeta_2}_{x\text{-}e/h,\pm,\mu,\nu_1}$ read:
	\begin{eqnarray}
	\mathcal{W}\,^{\zeta_1,\zeta_2}_{x\text{-}e/h,\pm,\mu,\nu_1} & = & \sum_{\nu_2,\textbf{\textit{k}}_1,\textbf{\textit{k}}_2} \hat{W}\,^{\zeta_1}_{x\text{-}e/h,\pm,\nu_2,\nu_1,\textbf{\textit{k}}_1,\textbf{\textit{k}}_2} \ f\,^{\zeta_2}_{e/h,\textbf{\textit{k}}_2} \sum_{\mu} \psi^R\,^{\zeta_1,\zeta_2}_{x\text{-}e/h,\pm,\mu,\nu_2,\textbf{\textit{k}}_1} , \\
	\hat{\mathcal{W}}\,^{\zeta_1,\zeta_2}_{x\text{-}e/h,\pm,\mu,\nu_1} & = & \frac{1\pm\delta^{\xi_1,\xi_2}_{s_1,s_2}}{2 \mathcal{A} \sum_{\textbf{\textit{k}}_1} f\,^{\zeta_2}_{e/h,\textbf{\textit{k}}_1}} \sum_{\nu_2,\textbf{\textit{k}}_2} \psi^L\,^{\zeta_1,\zeta_2}_{x\text{-}e/h,\pm,\mu,\nu_2,\textbf{\textit{k}}_2} \sum_{\nu_3,\textbf{\textit{k}}_3} \big(S\,^{\zeta_1}_{x\text{-}e/h,\pm}\big)^{-1}_{\nu_2,\nu_3,\textbf{\textit{k}}_2,\textbf{\textit{k}}_3} \sum_{\textbf{\textit{k}}_4} \big(\hat{W}\,^{\zeta_1}_{x\text{-}e/h,\pm,\nu_3,\nu_1,\textbf{\textit{k}}_3,\textbf{\textit{k}}_4}\big)^* f\,^{\zeta_2}_{e/h,\textbf{\textit{k}}_4} . \arxiv
	\end{eqnarray}	
	
	\twocolumngrid
%

\end{document}